\documentclass[12pt,preprint]{aastex}
\usepackage{amssymb}
\usepackage{amsmath}
\usepackage{xspace}
\usepackage{lscape}

\def\uf{\ensuremath{u_\mathrm{f}}\xspace}

\def\dx{\ensuremath{\mathrm{d}}}

\shorttitle{Chemistry in disks with grain evolution}
\shortauthors{Vasyunin et al.}

\begin{document}

\title{Impact of grain evolution on the chemical structure of protoplanetary disks}

\author{A.I. Vasyunin}
\affil{Max Planck Institute for Astronomy, K\"onigstuhl 17, D-69117
Heidelberg, Germany} \affil{Department of Physics, The Ohio State
University, 191 W.Woodruff Ave., Columbus OH, 43210,
USA\footnote{Moved in September
2010}}\email{vasyunin@mps.ohio-state.edu}
\author{D.S. Wiebe}
\affil{Institute of Astronomy of the Russian Academy of Sciences,
Pyatnitskaya str. 48, 119017 Moscow, Russia}\email{dwiebe@inasan.ru}
\author{T. Birnstiel}
\affil{Max Planck Institute for Astronomy, K\"onigstuhl 17,
D-69117 Heidelberg, Germany}\email{birnstiel@mpia.de}
\author{S. Zhukovska}
\affil{Max Planck Institute for Astronomy, K\"onigstuhl 17,
D-69117 Heidelberg, Germany}\email{zhukovska@mpia.de}
\author{T. Henning}
\affil{Max Planck Institute for Astronomy, K\"onigstuhl 17,
D-69117 Heidelberg, Germany}\email{henning@mpia.de}
\author{C.P. Dullemond}
\affil{Max Planck Institute for Astronomy, K\"onigstuhl 17,
D-69117 Heidelberg, Germany}\email{dullemon@mpia.de}

\begin{abstract}
We study the impact of dust evolution in a protoplanetary disk around a T~Tauri star
on the disk chemical composition. For the first time we utilize a comprehensive model of
dust evolution which includes growth, fragmentation and sedimentation.
Specific attention is paid to the influence of grain evolution on the penetration of the UV field
in the disk. A chemical model that includes a
comprehensive set of gas phase and grain surface chemical reactions is used to simulate
the chemical structure of the disk.

The main effect of the grain evolution on the disk chemical
composition comes from sedimentation, and, to a lesser degree, from
the reduction of the total grain surface area. The net effect of
grain growth is suppressed by the fragmentation process which
maintains a population of small grains, dominating the total grain
surface area. We consider three models of dust properties. In model
GS both growth and sedimentation are taken into account. In models
A5 and A4 all grains are assumed to have the same size ($10^{-5}$ cm
and $10^{-4}$ cm, respectively) with constant gas-to-dust mass ratio
of 100. Like in previous studies, the ``three-layer'' pattern (cold
midplane, warm molecular layer, and hot atmosphere) in the disk
chemical structure is preserved in all models, but shifted closer to
the midplane in models with increased grain size (GS and A4). Unlike
other similar studies, we find that in models GS and A4 column
densities of most gas-phase species are enhanced by 1--3 orders of
magnitude relative to those in a model with pristine dust (A5),
while column densities of their surface counterparts are decreased.
We show that column densities of certain species, like C$_{2}$H,
HC$_{2n+1}$N ($n=0-3$), H$_{2}$O and some other molecules, as well
as the C$_{2}$H$_{2}$/HCN abundance ratio, which are accessible with
Herschel and ALMA, can be used as observational tracers of early
stages of the grain evolution process in protoplanetary disks.

\end{abstract}


\section{Introduction}

The presence of numerous exoplanets and protoplanetary disks (PPD),
as well as our own existence, strongly suggest that planet
formation is ubiquitous in the Milky Way
\citep[e.g.,][]{51Peg,Mayor_Frei03,Udry_Santos07}. Still, it is one
of the most challenging problems for the modern astronomy to
understand how planets form. This topic covers a tremendous range of
micro- and macrophysics, and involves a wide variety of physical,
dynamical, and chemical processes.

According to the current paradigm, planets are assembled in
protoplanetary disks surrounding young stars, starting from the
coagulation of sub-micron dust grains into larger bodies
\citep[e.g.,][]{Natta_ea06, Henning08}. Circumstantial evidence indicates that
this process is quite effective and rapid. For example, large-scale IR
spectroscopic surveys of young stars in stellar clusters of various
ages tightly constrain typical dispersal timescales of circumstellar dust
disks of $\sim 5-10$~Myr, and mass accretion rates of $\dot{M}_{\rm
Sun} \sim 10^{-8}~M_\odot$~yr$^{-1}$
\citep{disc_fraction,SiciliaAguilar_ea06,Hernandez_ea08,Oliveira_09,
SiciliaAguilar_ea09, Fedele_ea10}. Inner, planet-forming zones of some older
disks are cleared of sub-micron-sized dust grains within just $\sim
1$~Myr \citep[e.g.,][]{Graham_ea07,Dutrey_ea08,Salyk_ea09, Thalmann_ea10}.
Furthermore, the analysis of the mineralogical and chemical composition
of unaltered chondritic meteorites shows that in the Solar Nebula
different types of meteorites have condensed and formed via grain
agglomeration within a period of $\sim 2-5$~Myr
\citep[e.g.,][]{Wasson_Kallemeyn88,Podosek_Cassen94,Thrane_ea06}.

However, observational data on grain growth are still scarce and
limited to the overall spectral energy distribution (SED) parameters. The presence of micron-sized
grains in disks has been inferred from mid-infrared spectroscopy
\citep[e.g.,][]{sil,Bouwman_ea08,Apai_ea04,vanBoekel_ea04,KesslerSilacci_ea06}.
A number of nearby protoplanetary disks has been imaged with the VLA
and Australia Telescope Compact Array (ATCA) at millimeter and
centimeter wavelengths, showing evidence of significant grain growth
up to at least pebble (cm) sizes
\citep[][]{Testi_ea03, Wilner_ea05,Rodmann_ea06,Natta_ea06,Cortes_ea09,Lommen_ea09}. On
the other hand, there can be other observables indicative of the
grain ensemble evolution.

In essence, grain evolution is controlled by collisions, leading to
coagulation and fragmentation of dust particles. For mm-sized particles the bouncing effect comes
into play, making the process more complicated and further growth less efficient~\citep{Zsom_ea10}.
Radial drift
and sedimentation to the disk midplane become important if the growth of grain particles proceeds further than about 1~cm.
Grains evolve faster in the dense disk interior ($\lesssim
10-20$~AU), while in the outer regions ($R\gtrsim100$~AU), the grain
ensemble may remain unchanged (ISM-like) for the entire life of the system
\citep[][]{Schmitt_ea97,Weidenschilling97,Ciesla07,Ormel_ea07,Brauer_ea08a,Birnstiel_ea09, Birnstiel_ea10}.
It is therefore reasonable to expect that the evolution of grain sizes and
spatial distribution is somehow reflected in the physical and
chemical structure of the disk.

In current models of disk structure it is assumed that the
temperature distribution within the disk, and thus its structure, is
controlled by the radiative transfer of energy generated due to
viscous dissipation and absorption of the stellar radiation. In the
disk atmosphere, photoelectric heating from dust grains becomes
important for energy balance \citep[e.g.,][]{GortiHollenbach09,
Woitke_ea09}. Dust opacities, which are changed by grain growth and
sedimentation, are the crucial element in this picture. Also,
because of dust evolution, stellar UV photons penetrate more easily
into the disk interior, heating it, and thus preventing rapid
freeze-out of molecules and enhancing photodestruction of gas-phase
chemical species
\citep[e.g.,][]{AikawaNomura06,Jonkheid_ea06,Jonkheid_ea07}. Local
variations in the dust-to-gas ratio, dust density and average grain
size may all affect the efficiency of gas-grain interactions, and
thus the efficiency of freeze-out and surface chemistry.

The problem of possible interrelations between dust evolution and chemistry has been addressed in the literature a number of times.
\citet{Jonkheid_ea04} studied the importance of dust settling on
the disk chemical evolution, assuming that the dust size distribution
remains the same along the disk radius. It was found that dust sedimentation has
a great impact on the gas
temperature in the disk and that dust and gas
temperatures are significantly different in the optically-thin disk region. On the other hand, sedimentation was
found to not strongly
influence the gas-phase chemistry in the model considered due to the
assumption that PAHs are still well-mixed with the gas and therefore
efficiently absorb stellar and interstellar UV photons. At the same
time, elevated temperatures of the upper disk layer in the model
with grain sedimentation lead to excitation of higher-lying
transitions.

\citet{AikawaNomura06} considered the effect of dust growth on
chemical abundances in a well-mixed disk around a low-mass star.
They modeled dust and gas temperatures in the disk
self-consistently and adopted a power-law grain size distribution.
The grain growth was simulated by raising the maximum grain
size up to $10\mu$m. An extended chemical network with gas-grain
interactions and surface processes was applied.
They found that the main effect of dust growth
is a shift of the molecular layer towards the disk midplane. At the same time,
  column densities
for many species were not affected by grain growth. Finally, they
identified a few sensitive tracers of grain growth in disks, such as
HCO$^+$, H$_3^+$, H$_2$D$^+$.

Later, \citet{Jonkheid_ea07} modeled chemistry and gas
temperature of evolving Herbig Ae disks with decreasing mass or dust
settling. They utilized a 2D radiative transfer code to model
gas/dust temperature and photodissociation rates. It was found
that the disk chemical structure strongly correlates with the disk mass:
while some photofragile molecules like HCN are abundant only in
massive disks, their fragments, like CN and CCH, become abundant in
low-mass disks. Furthermore, dust settling affects the gas temperature in the
disk atmosphere, and consequently abundances of disk atmosphere
tracers like O and C$^+$. They identified that line intensity ratios
of CO/$^{13}$CO, CO/HCO$^+$ and [OI] 63$\mu$m/14$6\mu$m can be used
to distinguish between disks undergoing grain settling and
photoevaporated disks.

In the present work, we move one step further and omit the simplifying
assumptions that grain evolution proceeds uniformly over the entire disk, or
that it can be described by artificially varying the
grain number density and upper grain size limit. The goal is
to isolate possible molecular tracers of grain evolution in the situation when
grain size distribution and dust-to-gas mass ratio vary smoothly in the disk,
as predicted by a dust evolution model. The model used includes coagulation, fragmentation, and sedimentation, assuming  equilibrium between turbulent stirring and
gravitational settling, as proposed in \citet{DullemondDominik04}. No bouncing effects are included. We focus on the combined effect of dust growth
and settling on the chemistry in a protoplanetary disk around a low mass
T~Tauri star. A large gas-grain chemical network with
surface reactions is utilized. This is the first paper in a series, where we intend to study various aspects of interrelations between grain evolution and chemistry.

The outline of the paper is the following. In section \ref{model_outline}, the setup we constructed to model the physical structure
of disk with evolving dust is described. Section \ref{chemical_model} contains a description of the detailed chemical model used to simulate the
chemical structure of the protoplanetary disk. In section \ref{results} we present modeling results. In paragraph \ref{dust_ev_in_disk} dust
evolution due to growth, fragmentation and sedimentation is discussed briefly. In paragraph \ref{uv_field_in_disk} change to the
UV field in the disk caused by evolution of dust properties are discussed. Paragraph \ref{disk_chemistry} is devoted to the analysis of the impact of
grain evolution on the chemistry of a protoplanetary disk. In section \ref{results}, we discuss whether or not the chemical composition of a
protoplanetary disk can be used as a tracer of grain evolution, and in which ways our model can be improved. Section \ref{conclusions} contains
the conclusions of this study.

\section{Description of the model}\label{model_outline}

A protoplanetary disk, even in a
simplified
description
is a complicated evolving system controlled by a large number of
interrelated physical and chemical processes. However, given the uncertainties and deficiencies in the physical
description of such a system, it would be impossible and impractical
to try to build a self-consistent dynamical model coupled to grain evolution and disk chemistry.
For this reason, we have
selected an approach which is by no means self-consistent, but is
relatively straightforward to implement and allows us to isolate the purely
chemical effect of the dust growth.

The disk chemical modeling in this study comprises of four
independent steps. First, we need a template of a disk physical
structure, i.e. gas density and temperature distributions. These
distributions are obtained from a 1+1D steady-state accretion disk
model. In the second step, the resulting thermal and density
structure is used as an input for the grain growth model. The third
step is to simulate the evolution of the dust grain ensemble in the
disk midplane due to grain coagulation, fragmentation and
sedimentation according to \cite{Birnstiel_ea09, Birnstiel_ea10}.
Third, the vertical distribution of dust is reconstructed in
stirring-mixing equilibrium based on the work by
\cite{DullemondDominik04}.

Finally, the grain parameters (size distribution at each disk location, as well as
dust-to-gas ratios) after 2~Myr of evolution are used in the simulation of the disk chemical
structure. These steps are described in more details in the
following subsections.

\subsection{Physical structure of the disk}

A popular formalism, which is used in many studies of disks
surrounding young stars, is based on the so-called $\alpha$-model of
\citet{ShakuraSunyaev73}. In this model it is assumed that the transport
of angular momentum in the disk is caused by turbulent
viscosity, with the viscosity coefficient $\nu$ expressed as $\alpha
H c_{\rm s}$, where $H$ is the disk semi-thickness at a given
radius, $c_{\rm s}$ is the isothermal sound speed, and $\alpha$ is
the viscosity parameter, which is usually assumed to be of the order
of $10^{-2}$ in protoplanetary disks.

As a first approximation for calculating the disk thermal structure of the disk, we assume thermal coupling between dust and gas. This assumption is well justified for optical depths $\tau \geq 1$, as was shown by the disk models with detail thermal balance \citep{AikawaNomura06, Jonkheid_ea06}. Sedimentation and coagulation of grains influence the disk temperature via a decrease in dust opacities \citep{Birnstiel_ea09}, which makes the atmosphere more transparent and hotter, and shifts the $\tau=1$ surface deeper into the disk. The investigation of the response of the disk thermal structure on dust evolution is a separate problem, which will be addressed by future papers in this series.

The disk structure is calculated with a simplified version of the
disk model described by \citet{dal1} and \citet{dal2}. Unlike
D'Alessio et al., in the calculation of gas temperature we take into
account only two heating sources, namely, viscous dissipation and
disk irradiation by the central star. Comparison of our
disk models with the D'Alessio et al. results shows that this
simplification does not significantly alter the derived disk physical
structure, and thus modeling of the disk chemistry is not affected.

The following equations are used to describe the disk vertical
structure:
\begin{equation}
{dP\over dz}=-g\rho,\label{structeq1}
\end{equation}
\begin{equation}
{dT\over dz}=-{3\kappa\rho F\over 4acT^3},\label{structeq2}
\end{equation}
\begin{equation}
{dF\over dz}=\frac94 \alpha P \Omega(R) + \Gamma_{\rm
irr}.\label{structeq3}
\end{equation}
Here $P$ is the gas pressure, $g$ is the acceleration due to gravity,
$\rho$ is the gas density, $T$ is the gas temperature, $F$ is the
flux of the thermal disk emission, $\Omega(R)$ is the disk (Keplerian) angular velocity at a distance $R$ from the central object. The first term on the right hand side of Eq. (\ref{structeq3}) gives the viscous heating of the disk, while
$\Gamma_{\rm irr}$ is the heating rate due to disk irradiation by
the central object.

The two-dimensional structure of the disk is modeled in the so-called
1+1D approximation. It is assumed that at each radius the disk is in
hydrostatic equilibrium (Eq.~\ref{structeq1}). Thus, the vertical structure of the disk
can be found from the solution of equations
(\ref{structeq1}--\ref{structeq3}) independently at each $R$. To
calculate the disk temperature at each vertical slice, we integrate
the transfer equation (\ref{structeq2}). The disk itself is
described by the following parameters:
\begin{itemize}
\item accretion rate $\dot{M}$;
\item viscous parameter $\alpha$;
\item temperature, radius, and mass of the central object $T_*, R_*,
M_*$ (radiation is assumed to be that of a black body);
\item inner and outer disk radii $R_{\rm in}$ and $R_{\rm out}$.
\end{itemize}

The outputs of the model are gas surface density distribution
$\Sigma_{gas}(R)$, and the 2D density and temperature structure of the gas.
The strengths of the UV and X-ray fields over the disk can then be
calculated using the information about the gas and dust properties in the disk.

Using this model, we simulated a DM Tau-like system. The system has
a disk with an outer radius of 550~AU, an inner radius of 0.03~AU,
an accretion rate $\dot{M}=2~\cdot~10^{-9}\,M_{\sun}$\,yr$^{-1}$,
and a viscosity parameter $\alpha = 0.01$. The disk is illuminated
by UV radiation from the central star with an intensity $\chi=500$
at $R=100$~AU in units of the mean UV field of \citet{Draine78} and
by the interstellar UV radiation. Two representations for the
stellar UV field are used. One is the scaled Draine field, and
another one is the representative observed spectrum as discussed by
\citet{Bergin_ea03}. The effect of these different radiation fields
will be discussed in section 3.1.

\begin{figure}
\includegraphics[width=0.55\textwidth]{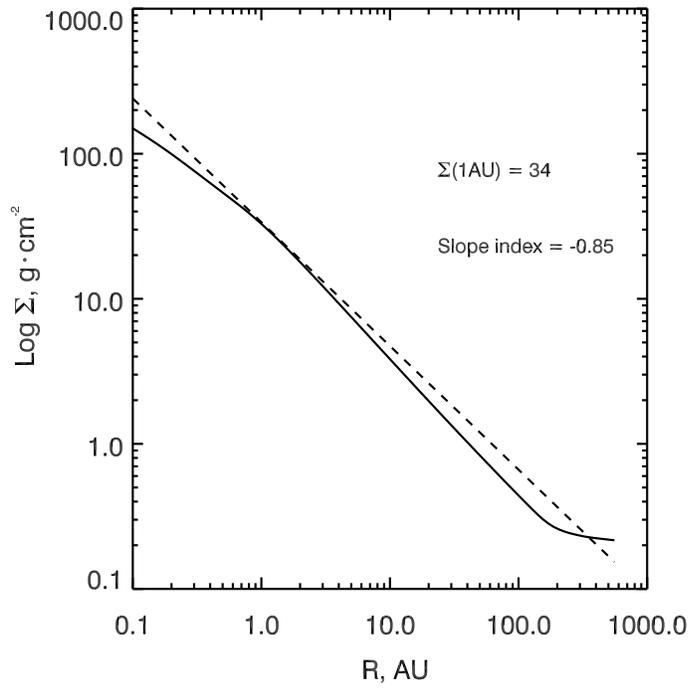}
\caption{The solid line shows the radial dependence of the gas
surface density in the adopted disk model. The dashed line
represents a power-law fit with the slope of --0.85.}
\label{surfdens05-full-paper}
\end{figure}

\begin{figure}
\includegraphics[width=0.45\textwidth,angle=270]{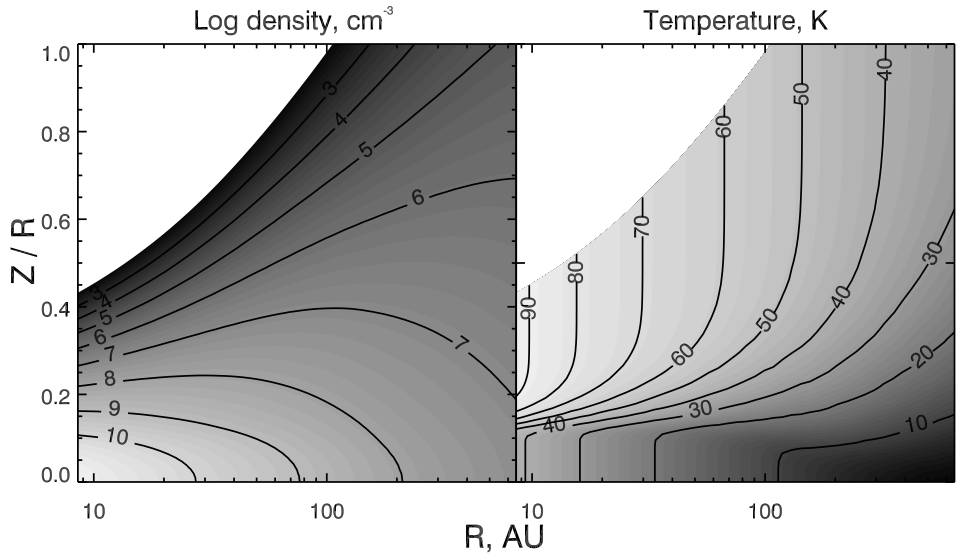}
\caption{Density and temperature distribution in the adopted disk model.}
\label{dens-temp-zr1}
\end{figure}

The star is assumed to have an X-ray luminosity of $10^{30}$ erg
s$^{-1}$ \citep{Glassgold_ea97}. X-rays induce the same reactions as
cosmic ray particles, so that X-ray ionization rate is in fact
simply added to cosmic ray ionization rate. The disk mass is
$M~=~0.055\,M_{\sun}$. The disk surface density profile and its
power-law approximation are depicted in
Fig.~\ref{surfdens05-full-paper}. The best-fit line has a slope of
0.84, which is somewhere in between of the Minimum Mass Solar Nebula
index (1.5) \citep{Weidenschilling77, Hayashi81} and the value
derived for PPDs from SED at millimeter wavelengths, ($\ge 0.5$)
\citep{AndrewsWilliams07}. The density and thermal structure of the
disk are shown in Fig.~\ref{dens-temp-zr1}. The main model
parameters are listed in Table~\ref{modpar}. We do not consider a
disk region beyond $R_{\rm out}$.

\subsection{Dust evolution in the disk}\label{dustmodel}

Several authors have constructed theoretical models for grain growth
in protoplanetary disks \citetext{e.g. \citealp{Nakagawa_ea81};
\citealp{Schmitt_ea97}; \citealp{Weidenschilling97}; \citealp{DullemondDominik05};
\citealp{Ormel_ea07}; \citealp{ZsomDullemond08};
\citealp{Brauer_ea08}
}. The main challenges of grain growth modeling are the following:
first, growth from $\sim$~0.01 $\mu$m grains to centimeter-sized
grains encompasses many orders of magnitude. A growth simulation must be able to conserve mass, since
large particles can grow by accreting large numbers of very small
grains. Second, particle size is not the only parameter determining the
properties of grains: porosity, composition and the various possible
outcomes of collisions further extend the parameter space \citep[see][]{Schmitt_ea97, Guettler_ea10}.

Here, we use a slightly modified version of the code presented in
\cite{Birnstiel_ea10} (see also \cite{Brauer_ea08}). This is a statistical, mass-conserving code which
implicitly solves the Smoluchowski equation, taking coagulation,
fragmentation and cratering into account. It is important to note
that we ignore radial drift of dust particles in the present study.

The mathematical formalism of the model is the following.
The number density distribution $n(m,r,z)$ is a function of mass
$m$, radius $r$, height $z$ above the midplane, and time. We do not consider grain porosity as an additional parameter. We define
the vertically integrated number density per mass $m$ as
\begin{equation}
N(m,r) \equiv \int_{-\infty}^{\infty} n(m,r,z) \, \dx z.
\label{eq:til:def_sigma}
\end{equation}

We assume that coagulation and fragmentation Kernels $K$ and $L$
(defined below) are independent of $z$. We can now describe the
time-evolution of this distribution by a vertically integrated
version of the Smoluchowski equation,
\begin{equation}
\begin{array}{lll}
\frac{\partial N(m)}{\partial t}
&=&+\int_{0}^{m/2} N(m') \: N(m-m') \: K(m',m-m')\:\mathrm{d} m'\\
\\
&&-\int_{0}^{\infty} N(m') \: N(m) \: K(m,m')\: \mathrm{d} m'\\
\\
&&+\frac{1}{2} \iint_{0}^{\infty} N(m') \: N(m'') \: L(m',m'') \: S(m,m',m'')\: \mathrm{d} m'\: \mathrm{d} m''\\
\\
&&-\int_{0}^{\infty} N(m') \: N(m) \: L(m,m')\: \mathrm{d} m',\\
\end{array}
\label{eq:til:smoluchowski}
\end{equation}
where the radial dependence has been omitted, since we treat each radius
independently, neglecting radial movement of dust. The right-hand side
terms of Eq.~\ref{eq:til:smoluchowski} (from top to bottom)
correspond to gain and loss by coagulation and gain and loss through
fragmentation.

The coagulation kernel $K(m,m')$ and the fragmentation kernel
$L(m,m')$ are then given by,
\begin{equation}
\begin{split}
K(m,m') &= \frac{1}{\sqrt{2\pi (h(m)^2 + h(m')^2)}} \cdot p_\text{c} \cdot \sigma(m,m') \cdot \Delta u(m,m')\\
L(m,m') &= \frac{1}{\sqrt{2\pi (h(m)^2 + h(m')^2)}} \cdot p_\text{f}
\cdot \sigma(m,m') \cdot \Delta u(m,m'),
\end{split}
\end{equation}
where $p_\text{c}$ and $p_\text{f}$ are the coagulation and
fragmentation probabilities, respectively, which have a sum of
unity, $\Delta u(m,m')$ is the relative particle velocity, and
$\sigma(m,m')$ is the sum of their geometrical cross sections. In
this work, we consider Brownian motion, vertical settling
\citep[see][]{Brauer_ea08}, and turbulent motion
\citep[see][]{OrmelCuzzi07} as physical effects that produce the
relative particle velocities.

Particles colliding with a relative velocity higher than the
critical velocity \uf are assumed to  fragment into a power-law
size distribution of fragments \citep[i.e., $S(m,m',m'')\propto
m^{-1.83}$, see][]{Brauer_ea08} if the particle masses differ by
less than one order of magnitude. Otherwise, the smaller body is
assumed to excavate mass from the larger one by cratering, where the
amount of excavated mass equals the mass of the smaller body. The
fragmentation velocity \uf is
taken to be 10~m/s \citep{Birnstiel_ea09}.

This model is related to a model of gas disk structure through two
input parameters: radial midplane temperature distribution $T_{\rm
mid}(R)$ and the radial gas surface density distribution
$\Sigma_{\rm gas}(R)$. A more comprehensive description of the
physics of coagulation/fragmentation and of its numerical
implementation can be found in \citet{Brauer_ea08} and
\citet{Birnstiel_ea10}.

The numerical model described above provides us with surface
densities of grains of different masses. To calculate the vertical
distribution $\rho_i(R,z)$ of grains, we assume that it is controlled
by the equilibrium between gravitational settling and turbulent
stirring, as proposed by \citet{DullemondDominik04}. Vertical scale heights
are computed separately for each size bin.

\subsection{Average grain size for the chemical model}

An output of the dust evolution model is the grain size distribution
at each disk location. In principle, this distribution can be
incorporated into the chemical model by considering surface
reactions and gas-dust interactions for each size bin separately.
However this would make the model too complicated, and we elect to
use a single grain size $\overline{a}(r,z)$ in the chemical model,
which is computed from the equilibrium local grain size distribution
in each disk location. The averaging is performed in a way which
preserves the total surface area and the total dust mass. For a
grain size distribution $f(a,r,z)$, where $f(a,r,z)da$ is the number
of grains with radii between $a$ and $a+da$ in the disk point with
circumstellar radius $r$ and height above midplane $z$, this means
that
\begin{equation}\label{aeq1}
\frac{4}{3}\pi\overline{a}^3(r,z)\overline{n} =
\frac{4}{3}\pi\int_{a_{min}}^{a^{max}}{f(a,r,z)}a^{3}da ,
\end{equation}
\begin{equation}\label{aeq2}
4\pi\overline{a}^2(r,z)\overline{n} =
4\pi\int_{a_{min}}^{a^{max}}{f(a,r,z)}a^{2}da,
\end{equation}
where $\overline{n}$ is the total number density of
``representative'' equal-sized particles. By dividing Eq. (\ref{aeq1})
by Eq. (\ref{aeq2}) one obtains
  \begin{equation}
  \overline{a}(r,z) =
  \frac{\int_{a_{min}}^{a^{max}}{f(a,r,z)}a^{3}da}{\int_{a_{min}}^{a^{max}}{f(a,r,z)}a^{2}da}.
  \label{meanrad0}
  \end{equation}

An ensemble of grains with the mean radius $\overline{a}(r,z)$ has
the same mass and total surface area as the original ensemble. We
use $\overline{a}(r,z)$ exclusively in our astrochemical model
described below when dealing with gas-grain interactions and surface
chemical processes.

It is important to note that the way we calculate the average grain
size is different from those usually used in works on grain growth
(e.g., \citet{Brauer_ea08, Birnstiel_ea10} etc.). In such works, the
averaging is mass-weighted, i.e., emphasis is put on the size of a
small fraction of the most massive grains. In this paper, on the
other hand, averaging is surface-weighted, i.e. emphasis is put on a
population of small grains which dominate the grain surface area as
well as the total number of grains. In other words, if simulations
of grain growth would give us a distribution of grains in which a
small number of big boulders co-exists with a big number of tiny
grains, the average grain size of such a distribution calculated
with expression (\ref{meanrad0}) will be close to the size of small
grains, contrary to the ``usual'' definition of average grain size.

\section{Chemical model of the disk}\label{chemical_model}

In principle, simultaneous modeling of the disk chemical evolution,
together with grain evolution, requires a non-local approach.
Grains not only grow, but also move within the disk due to
sedimentation and radial drift \citep[e.g.]{Brauer_ea08}, bringing
with them molecules in icy mantles. However, a coupled chemo-dynamical
disk model with grain growth would be computationally expensive, and
difficult to analyze from the chemical point of view, so in the present study
we choose to neglect transport effects. Thus, we model a chemical
structure in the disk, taking dust properties and physical
conditions from the previously-described disk model, and assuming
that these conditions do not change over the timespan of
chemical evolution considered. This timespan is taken to be 2~Myr, but the
particular choice does not significantly affect our conclusions, as the
abundances of most species are equilibrated very fast.

In this study, we utilize the chemical model described in
\citet{Vasyunin_ea08}. Initial elemental abundances are taken from
Table~1 in \cite{Vasyunin_ea08}. Compared with the original chemical
model, we made several modifications. First, dissociative
recombination of ions on grain particles is included according to
\cite{UmebayashiNakano80} and \cite{Semenov_ea04}. Second, for
gas-grain chemical interactions and grain surface reactions, we used
variable dust-to-gas mass ratios and the average grain size
calculated at each disk point using data from the grain growth
model. For grain surface reactions, we utilized model~H from
\citet{Vasyunin_ea09}. In this model, only thermal hopping is a
source of mobility of surface species, and a high
diffusion-to-desorption energy ratio of 0.77 is adopted for all
species. Under these conditions, stochastic effects in grain surface
chemistry are negligible and classical rate equations may be safely
used\citep{Vasyunin_ea09, Garrod_ea09}. In our model molecular
hydrogen is formed via the surface reaction
H~+~H~$\rightarrow$~H$_{2}$. The rate of this reaction, as well as
of other surface reactions, is calculated according to the formalism
described by \cite{Hasegawa_ea92}. The rates are inversely
proportional to the surface area of an individual dust grain.

The cosmic ray ionization rate has been adopted according to Eq.~19
from \cite{Sano_ea00}. The unattenuated cosmic ray ionization rate
was taken to be $1.3\cdot10^{-17}$~s$^{-1}$. The final and most
important change is the detailed treatment of photoprocesses which
take into account the shape of the UV spectrum of the central star,
and its attenuation in the disk due to absorption of the UV photons
by dust grains (see Section \ref{pr}). Below we discuss in detail
the treatment of these photoprocesses.

\subsection{Photoreactions}\label{pr}

Photoreactions represent the crucial element of chemical networks
designed for the upper disk atmosphere, and it is this element that
is most affected by the dust evolution. In the conventional
approach, photoreaction rates are estimated as
\begin{equation}
k^{\rm ph}=\chi k_0 \exp(-\gamma A_{\rm V}), \label{phstd}
\end{equation}
where $k_0$ is the reaction rate for the unshielded interstellar UV
radiation field, $\chi$ is the UV intensity scaling factor, and
$\gamma$ is a parameter used to account for the different dust
extinction in the UV and visual wavelengths. Generally speaking,
$A_{\rm V}$ estimated from the gas column density cannot be used in
this case, because it is based on a particular choice of dust
properties (opacities and dust-to-gas mass ratio), and on the
assumption that these properties do not vary in the medium. In our
study dust, parameters are position-dependent, so we have to be more
careful. To compute rates of photodissociation and photoionization
one uses the general expression:
\begin{equation}
k^{\rm ph} (r,z)= 4\pi \int {I_\nu(r,z)\over h\nu} \sigma_\nu d\nu,
\end{equation}
where the mean UV field intensity $I_\nu(r,z)$ is obtained as a
solution of the radiation transfer problem. In this study we adopt a
simplified approach, in which $I_\nu(r,z)$ is given by the equation
\begin{equation}
I_\nu(r,z)=I^{\rm st}_\nu(r,z)\exp(-\tau^{\rm st}_\nu(r,z))+I^{\rm
D}_\nu\exp(-\tau^{\rm IS}_\nu(r,z)),
\end{equation}
where $I^{\rm st}_\nu(r,z)$ is the unattenuated diluted stellar
radiation field, $I^{\rm D}_\nu$ is the so called Draine radiation
field \citep{Draine78}. We investigated two representations for the
stellar radiation field. In the first representation the observed
spectrum for TW Hya is used, scaled in such a way that the
integrated UV intensity is about 500 in units of the Draine field at
100 AU from the star \citep{Bergin_ea03}. A combination of FUSE
\citep{TWHya_FUSE} and HST \citep{starcat} data is utilized, and the
resulting spectrum is shown in Fig.~\ref{spectrum}. Essentially,
this spectrum is very close to the one used in \citet{Bergin_ea03}.
In another representation, the stellar radiation is assumed to be
given by the scaled Draine interstellar field, with the same
normalization. Comparison of results for the two representations
shows that both spectra lead to the same general conclusions about
the chemical structure of the disk both with evolved and unevolved
dust. This is consistent with findings of \cite{vanZadelhoff_ea03}.
These authors compared molecular distributions and column densities
for the cases when the UV spectrum is given either by the scaled
ISFR spectrum or the smoothed TW Hya spectrum and found no
significant differences (their models A and B). In our case, the
relative similarity of results for the two representations is caused
by effective attenuation of the stellar radiation in the disk, so
that the radiation field is in both cases dominated by the
interstellar field everywhere in the disk, except for the uppermost
layers. The situation would have been different if scattering were
present in our model. When the scaled ISRF is used to represent the
stellar radiation, scattering is in some sense equivalent to just a
somewhat different scaling factor for the vertically penetrating
radiation. However, when a more realistic stellar spectrum is used,
like the L$\alpha$-dominated TW Hya spectrum, scattered stellar
radiation would noticeable change photoreaction rates for species
with large cross-sections in the L$\alpha$ range
\citep{Bergin_ea03}. Note that \cite{vanZadelhoff_ea03} did find a
significant dependence on the spectrum of the radiation field in a
more general sense, that is, for the case when no UV excess is taken
into account at all (their model C).

Also, \cite{vanZadelhoff_ea03} investigated how different approaches
to the radiation transfer, namely, full 2D and simplified
calculations, such as described below, impact the chemical
composition of a protoplanetary disk. They found that column
densities of most species are not sensitive to the chosen approach.
Important exceptions are the molecules CN, C$_{2}$H, CS and HCN, but
only in the inner disk. Their column densities do significantly
depend on the adopted UV radiation transfer model. However, in our
model, the strength of the stellar UV field is more than an order of
magnitude weaker than that considered by \cite{vanZadelhoff_ea03}.

The optical depth $\tau^{\rm st}_\nu(r,z)$ is computed along the ray between
the star and the current location, so that
\begin{equation}
\tau^{\rm st}_\nu=\int \kappa_\nu(r,z)\rho(r,z) ds.
\end{equation}
Dust sedimentation is taken into account in the dust density
$\rho(r,z)$. The absorption coefficient $\kappa_\nu(r,z)$ per unit dust
mass is calculated at each location along the ray as
\begin{equation}
\kappa_\nu(r,z)=C\int {\rm d}a f(a,r,z)Q(a,\nu) \pi a^2,
\end{equation}
where the normalization coefficient $C$ is defined by the condition
\begin{equation}\label{norm_eq}
1=C\int {\rm d}a f(a,r,z)\frac43 \pi a^3\rho_{\rm d}.
\end{equation}
Absorption efficiency factors $Q(a,\nu)$ are taken from
\citet{DraineLee84} and \citet{WD2001}. For the purpose of opacity
computation, we assume that dust consists of silicate and carbon
(graphite) components contributing, correspondingly, 70\% and 30\%
to the total dust mass density \citep{WD2001}. In Eq.~\ref{norm_eq}
$\rho_{\rm d}$ is the density of dust material, taken to be 3 g
cm$^{-3}$ for silicate grains and 2.24 g cm$^{-3}$ for graphite
grains. In the expressions above, $f(a,r,z)$ is the local evolved
grain size distribution that comes from the dust growth and
sedimentation model. Considering coagulation and fragmentation, we
do not distinguish between the two materials as experimental data
show a similar dust evolution behavior for both types of grains (J.
Blum, private communication).

The optical depth $\tau^{\rm IS}_\nu(r,z)$ is computed in the same way,
but in vertical direction from the disk surface
to the height $z$.

The reaction cross sections are taken from \citet{vanDishoeckFaDi}
and supplemented with the additional data for the Ly$\alpha$
wavelength, with the exception of the photodissociation cross
sections for NO, HCN, NO$_2$, SO$_2$, OCS, NH$_3$, CH$_4$,
H$_2$O$_2$,  and C$_2$H$_2$. For these molecules, monochromatic
photodissociation cross sections from the
AMOP\footnote{http://amop.space.swri.edu/} database have been used
in our study.

To compute the rates of photoreactions for which
wavelength-dependent cross-sections are not available, we use the
standard expression (\ref{phstd}), however with the modified
procedure for computing $A_{\rm V}$. The conventional expression for
optical extinction
\begin{equation}
A_{\rm V}=N_{\rm H}/1.59\cdot10^{21}\,{\rm cm}^{-2}
\end{equation}
is valid for dust which is well-mixed with gas and consists of
single-sized grains ($10^{-5}$ cm) and a dust-to-gas mass ratio
0.01, but it can easily be re-normalized for other values of these
parameters. We integrate $A_{\rm V}$ along the ray between the star
and the considered location (or in a vertical direction for
interstellar radiation), taking into account the variable
dust-to-gas ratio along the ray, and a variable average grain size
$\overline{a}(r,z)$.

Self-shielding for H$_2$ photodissociation is computed using the
\cite{DraineBertoldi96} formalism, with the modified $A_{\rm V}$
value used to account for dust attenuation. The self- and mutual
shielding for CO photodissociation is computed with data from
\cite{Lee_ea96}. Their dust attenuation is not included in the
resultant shielding value, as it is replaced by our optical depths.
To compute self-shielding factors, one would need to know the CO
column density along the ray, which in the chemical study would
entail iterations. We adopted a much simpler approach, assuming that
CO self-shielding is proportional to the overall column density with
an average abundance of 10$^{-6}$. To check whether this
simplification affects our conclusion, we run two models, bracketing
possible self-shielding values. In one of the models, maximum
self-shielding is assumed which would correspond to all C atoms
being in CO molecules. In the other model, CO self-shielding is
assumed to be zero. Comparison of these two models showed that
various treatments of self-shielding do not change our main
conclusions while they do change some numbers, so we believe that
the use of this simplification is acceptable. Anyway, species that
turned out to be sensitive to the CO self-shielding treatment (CS
and H$_2$CO) are not included in Tables \ref{tab:sens_spec} and
\ref{tab:NOTsens_spec}.

In the model, several important simplifications are made. First, we
assume that dust and gas temperatures are equal everywhere in the
disk. Next, we neglect transport processes, such as radial and
vertical mixing, and their influence on molecular abundances.
Finally, when simulating grain-surface chemistry, we utilize the
single average grain size instead of the real grain size
distribution. Impact of these simplifying assumptions on the results
of this study is discussed in Section \ref{discussion}.

\section{Results}\label{results}

\subsection{Dust evolution in the disk}\label{dust_ev_in_disk}
\begin{figure}
\includegraphics[width=0.9\textwidth]{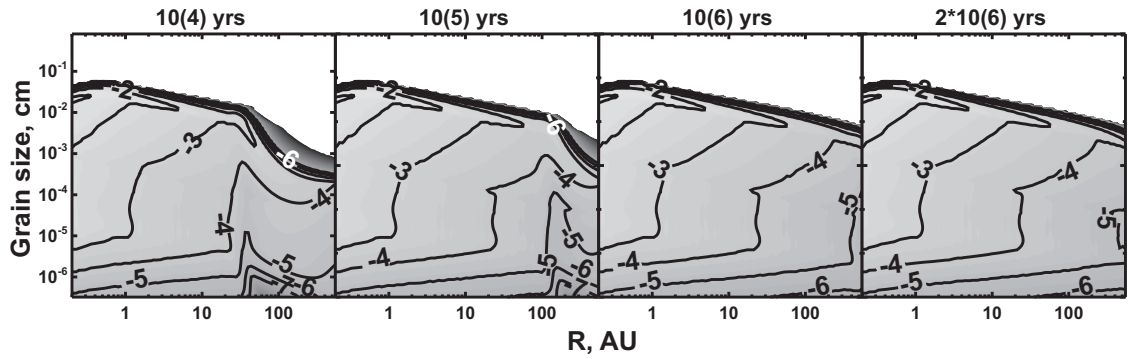}
\caption{Evolution of surface density of growing dust in model GS.
Contours show surface density in g\,cm$^{-2}$ for dust grains of a
certain size at a certain radius, so that, e.g., surface density of
0.1\,$\mu$m grains at 1~AU is about $10^{-3}$~ g\,cm$^{-2}$ after
$10^4$ years of evolution (leftmost panel). Dust growth is faster in
the dense inner part of the disk. Steady-state grain size
distribution is reached there between 10$^{4}$ and 10$^{5}$ years.
In the outer disk, density is lower and steady-state distribution is
reached between 10$^{6}$ and 2$\cdot$10$^{6}$ years.}
\label{dust_surf_dens}
\end{figure}

\begin{figure}
\includegraphics[width=0.5\textwidth,angle=0]{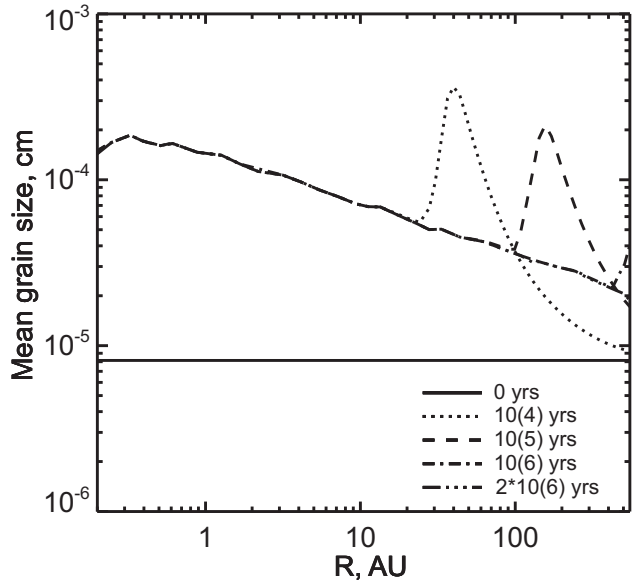}
\caption{Radial distribution  of
the vertically-integrated average grain radius at different
time moments. Averaging is performed in a way that preserves the total grain mass and surface area.}
\label{grainsize-mrn}
\end{figure}

Using the model of grain growth and fragmentation described in
Section \ref{dustmodel}, we calculated the evolution of surface
densities for grains of different sizes, starting from the standard
MRN grain size distribution at $t=0$. \citep{Mathis_ea77}. Snapshots
of surface density as a function of grain size for four time moments
are shown in Fig.~\ref{dust_surf_dens}. The corresponding vertically
integrated average grain size calculated according to Eq.
(\ref{meanrad0}) using vertically integrated number densities from
Eq. (\ref{eq:til:def_sigma}) is shown in Fig.~\ref{grainsize-mrn}.

The two peaks in Fig.~\ref{grainsize-mrn} appear because small
grains typically grow together in a more or less uniform bump moving
to larger sizes. So, the whole distribution shifts up in size until
grains reach the fragmentation barrier. Once that happens, most of
them fragment, and the small sizes get re-populated. The average
grain size drops again, and an equilibrium between growth and
fragmentation is established.

Since the disk density is higher closer to the star, grain growth
proceeds faster there (in our model we do not consider the radial
drift of grains). One can see that at $R \le 3$0~AU a steady-state
in the grain surface density distribution is reached within less
than $10^{4}$~yrs. However, in the outer disk ($R \ge 100$~AU) grain
surface density evolves much slower. A steady-state distribution is
barely reached there within $10^{6}$~yrs. In order to ensure that
grain parameters are equilibrated at all radii, we ran the grain
evolution model for 2~Myr.

It is important to point out that the highest surface density
(Fig.~\ref{dust_surf_dens}) corresponds to grains which are bigger
than the average particle size, as shown in Fig.~\ref{grainsize-mrn}.
It is in these large grains that most of the dust mass is contained,
while small grains dominate the dust surface area. Therefore, the
presence of small grains in the disk mediates the net effect of
grain growth from the point of view of gas-grain interactions, as
they depend primarily on the available grain surface.

In general, the grain growth is quite modest in the modeled disk.
The largest increase in the surface averaged grain size is achieved
in the inner disk midplane ($10~\le~R~\le~100$~AU) with a mean grain
size of 10$^{-4}$~cm (1 $\mu$m). In the outer disk region ($R \ge
100$~AU) the average grain radius increases relative to the initial
value by a factor of only 3--5.

\begin{figure}
\includegraphics[width=0.55\textwidth]{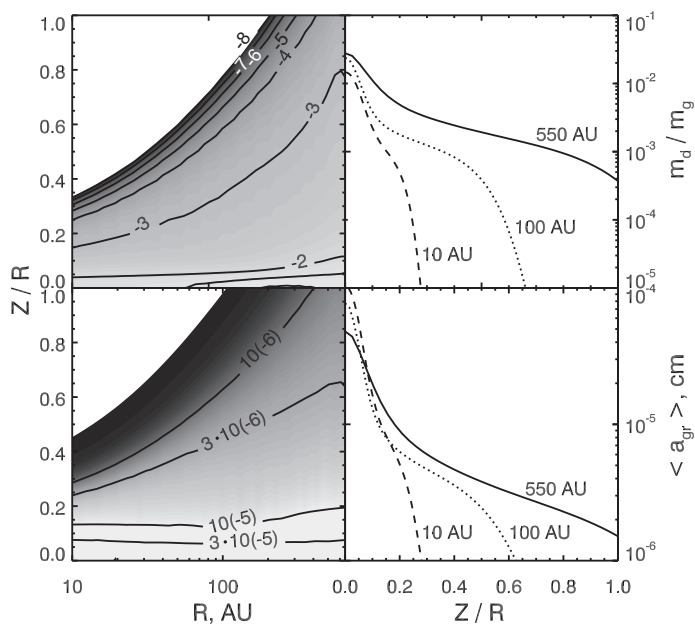}
\caption{Dust-to-gas mass ratio (upper row) and an average grain radius
(bottom row) in model of the disk with grain growth and sedimentation. At the left panels, 2D distributions over the disk are shown
while at the right panels vertical distributions at three representative radii are plotted.
Large grains that contain
most of dust mass sediment to the midplane more efficiently than
small ones. This reduces the dust-to-gas mass ratio above the
midplane and decreases the average grain size in upper layers of the
disk.} \label{mdmg-zr1}
\end{figure}

In contrast, differential dust settling seems to be much more
significant. 2D distributions of the total dust-to-gas mass
ratio and average grain size are shown in Fig.~\ref{mdmg-zr1}.
Dust settling reduces the dust-to-gas
mass ratio in disk regions above the midplane. Everywhere above $Z/R
\sim 0.1$ this ratio is significantly below the
canonical value of 0.01. In the chemically rich intermediate layer it
varies between 0.001 and 0.01, while in the disk atmosphere it drops
down to $10^{-4}-10^{-5}$. At the same time, in the midplane the dust-to-gas mass ratio can reach a value of
0.02. Similarly, the average
grain radius decreases above the midplane, though its
change is not that dramatic. In the disk atmosphere, the average grain radius
is only $10^{-6}$~cm (0.01 $\mu$m), because large grains efficiently settle down to the midplane.

\begin{figure}
\includegraphics[width=0.45\textwidth]{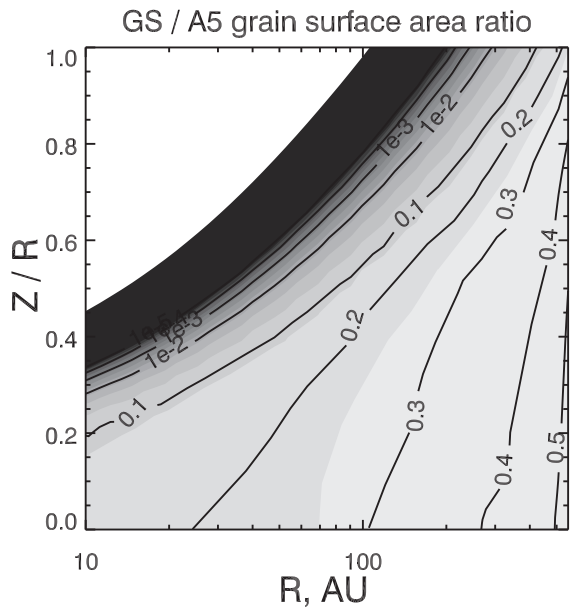}
\caption{Ratio of grain surface areas in models with-- and without grain evolution. Grain evolution reduces the total grain surface area.
This effect is more significant in the inner disk because grain growth is more efficient there, and also above the midplane because of
grain sedimentation to the midplane.}
\label{totsurfarea-new2old-zr1}
\end{figure}

Summarizing both the vertically-averaged and 2D parameters of the evolved dust, we note
that the overall effect of the grain growth and sedimentation in our disk
model is the global decrease of the total surface area of dust
grains (see Fig.~\ref{totsurfarea-new2old-zr1}). This is less pronounced in the outer part of the disk ($R \ge 100$~AU), corresponding to a factor
of only 2--3. In the inner disk region ($R \le 100$~AU) the decrease
of the dust surface area is more prominent, i.e., up to one order of
magnitude. The ratio of evolved to original grain surface area has a
steeper radial dependence than variations in the vertical direction. This is because the decrease
in total dust surface area above the midplane, caused by the dust settling, is partly compensated
by the presence of small particles dominating the surface area.

\subsection{UV field in the disk}\label{uv_field_in_disk}

For the radiation field in the disk, we consider three basic models. In model A5, all grains are assumed to have a single size of $10^{-5}$~cm. This is the
model of choice in most astrochemical computations. Model~A4, in which all grains have the same radius of  $10^{-4}$~cm, can be
considered a rough representation of grain growth (without
sedimentation). Finally, in model~GS both grain growth and sedimentation
are taken into account fully. For brevity we will call the GS and A4 models ``evolved'', while the A5~model
is referred to as the ``standard'' model.

In all models, the disk is illuminated by both stellar and
interstellar UV radiation. For disk chemistry, the strength and
shape of the UV field is of primary importance, as it controls
the rates of photoionization and photodissociation, as well as the efficiency of
the photodesorption of grain mantles \citep[e.g.,][]{Bourdon_ea82,
Westley_ea95, vanZadelhoff_ea03, Oeberg_ea07}. Due to the presence
of dust, the UV field is attenuated as it penetrates deep into the
disk. As mentioned in the introduction, the attenuation of starlight in the disk atmosphere
can also be caused by PAH particles. IR observations of protoplanetary
disks show that PAHs are an abundant component of some disks around Herbig~Ae stars \citep[e.g.,][]{Acke_ea10},
but they are usually not seen in the spectra of T~Tauri stars \citep[][]{Geers_ea06}. Therefore, for
simplicity, we neglect their possible contribution to the opacity and heating of the disk.

\begin{figure}
\includegraphics[width=0.9\textwidth]{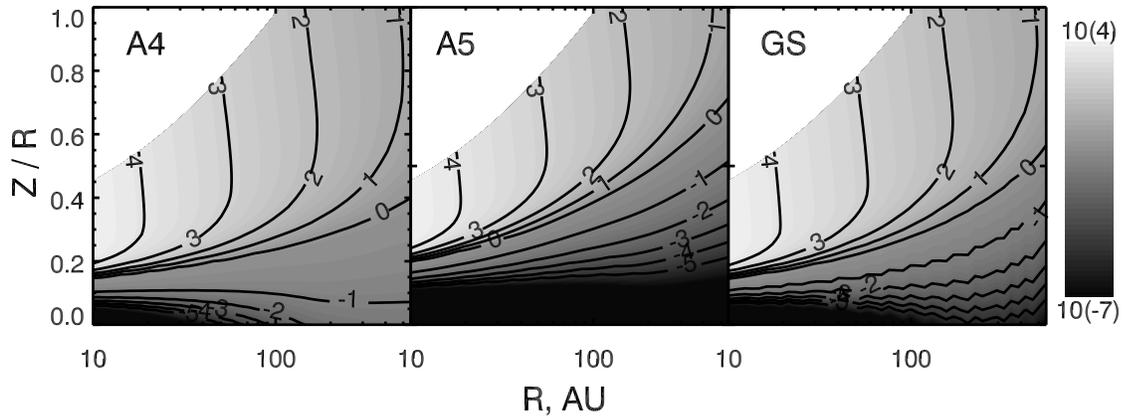}
\caption{Logarithm of the UV field intensity in the disk for models A5, A4 and
GS measured in units of the interstellar UV radiation field. Changes in grain properties strongly affect the amount of UV radiation penetrating into the disk. While A5 model has a dark
midplane area at all radii, the model A4 disk is relatively transparent to the UV radiation beyond 100~AU. In the model GS the extent of the
dark region is smaller than in model A5 (for R$\le$100~AU) and there is a ``dim'' rather than a ``dark'' midplane beyond 100~AU.} \label{UV4}
\end{figure}

\begin{figure}
\includegraphics[width=0.75\textwidth]{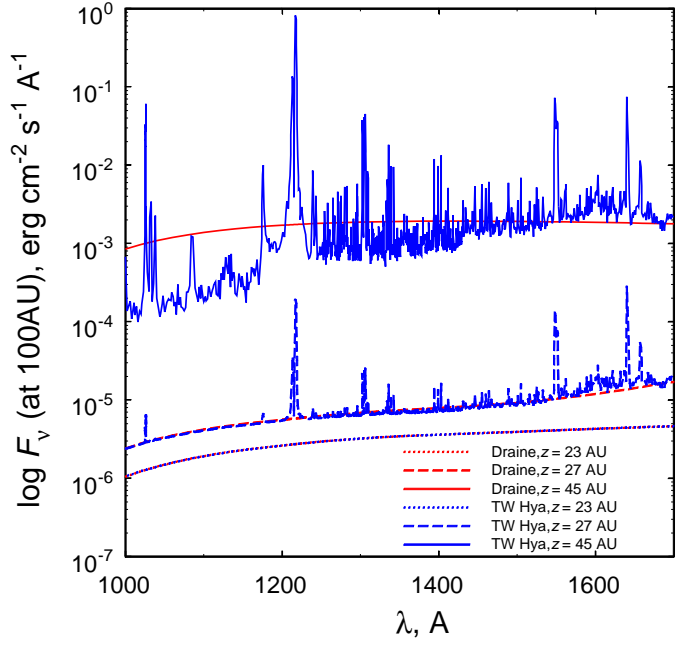}
\caption{Spectrum of the UV radiation in the disk at different
heights above the midplane at $R=100$~AU in model GS. Closer to the
midplane, the spectrum shape is similar to that of the ISRF because
the stellar component is attenuated faster than the interstellar
one. For comparison, spectrum of the scaled-up Draine field is also
shown.} \label{spectrum}
\end{figure}

In Fig.~\ref{UV4}, the distribution of the integrated UV field strength
over the disk is shown in logarithmic scale. Contours are labeled in units of the interstellar UV radiation
field \citep[][]{Draine78}. In the evolved models the disk becomes more transparent to the incident UV
radiation. While in model~A5 (Fig.~\ref{UV4}, middle panel) one can clearly see
the dark zone at the disk midplane (up to $Z/R=0.15$), this region is significantly smaller in models GS
and A4. On the other hand, in evolved models, the midplane can be divided into two parts. The inner part at
$R\le 100$~AU remains as opaque to UV radiation as in the standard model,
while the outer midplane can hardly be called ``dark'', because of the smaller opacities in models GS and A4.
In model~A4, the UV strength in this region is $\sim~10^{5}$ times
higher than in model~A5. Thus, the outer disk in model A4 consists of only two
layers: the atmosphere and the moderately UV-illuminated
midplane (Fig.~\ref{UV4}, left panel). In the GS model the outer disk is not that transparent, even though the inner dark region of the
midplane is still narrower than in model~A5. The UV
strength in the outer disk region is higher than in model~A5 by a
factor of $10-10^{3}$, and is only a small fraction of the UV
strength in the intermediate layer. Therefore, the three-layered disk
chemical structure is preserved in the GS model but the midplane is only ``dim'' rather than ``dark'' (Fig.~\ref{UV4}, right panel).

The UV spectrum at various
heights above the midplane for a disk radius of $R=100$~AU is shown in Fig.~\ref{spectrum} .
For illustrative purposes, we also schematically include the Ly$\alpha$ feature. As expected,
the dust absorption is stronger at shorter
wavelengths. It is important to note that Ly$\alpha$ feature
``disappears'' in the densest part of the disk, due to the strong
absorption and the absence of any efficient scattering
mechanisms for UV photons in our model. Hence, we expect only
minor differences in column densities of molecules in the current version of our model with and
without Ly$\alpha$, because the major contribution to column densities comes from the densest
parts of the disk.
In future works, we plan to treat
the scattering of Ly$\alpha$ photons
\citep[see][]{Bergin_ea03}.

Another factor that can affect the UV field in the upper disk atmosphere is the gas-phase opacity. Recently,
\cite{BethellBergin09} noted that water and OH molecules can be a source of significant absorption in the UV band. However, to take this effect into account properly, one would need to model chemical and physical disk structure self-consistently.

\subsection{Chemical structure of the disk}\label{disk_chemistry}

We now want to discuss which effects -- grain growth or sedimentation -- are dominating changes in
column densities of the various gas phase species.
Note, that in all three basic models detailed cross-sections
are used, to calculate photoreaction rates whenever possible (see section \ref{pr}).

\subsubsection{Inner disk: models GS and A5}
First, we compare models GS and A5. The overall trend for these two
models is that the column densities in the inner disk are greater
for most species (like CO, CO$_2$, NH$_3$) in the GS model than in
the A5 model, sometimes by orders of magnitude. Of all the gas-phase
species, only 10\% have greater column denisities in the A5 model at
$R=10$~AU (like S$^+$, HNO, CP). This result may seem
counterintuitive, as the disk which is transparent to UV radiation
should represent a more hostile environment for molecules. However,
at the same time the growth of dust particles decreases their total
surface area, thus lessening the effectiveness of mantle formation
and surface chemical processing. Also, the more intense UV field in
the upper disk causes faster photodesorption, shifting the balance
between sticking and desorption toward higher gas-phase abundances.

\begin{figure}
\includegraphics[width=0.3\textwidth,angle=90]{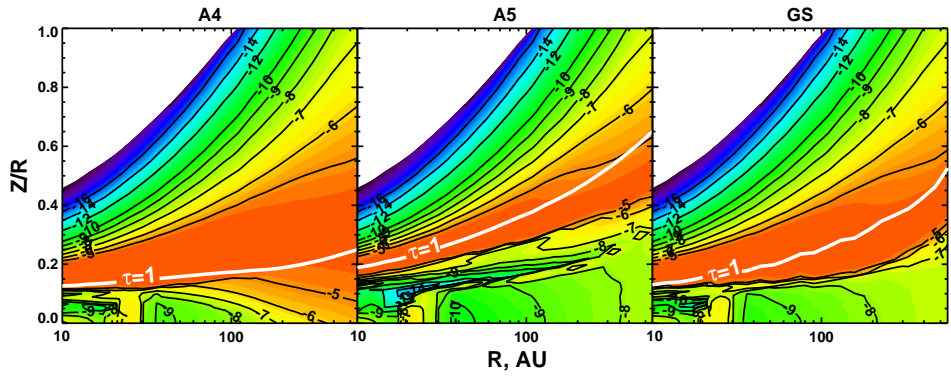}
\caption{2D distributions of CO abundance of the species relative to
H$_{2}$ in models A4, A5 and GS respectively.}\label{CO-2D}
\end{figure}

The most dramatic example of this difference is H$_2$CS. The column
densities of this molecule at 10~AU in models GS and A5 differ by
almost 6 orders of magnitude. However, there are also less exotic
examples represented by basic carbon compounds, also at 10~AU, where
the effects of dust evolution are stronger. Both CO (see Fig.
\ref{CO-2D}) and CO$_2$ are concentrated primarily in the warm
molecular layer. The peak gas-phase abundances of CO are nearly the
same in models GS and A5, as they are essentially equal to the total
carbon abundance. However, the location of the peak of the relative
abundance of CO in the GS model is shifted downward to the denser
disk region causing the CO column density to increase. The peak of
the CO$_2$ abundance is also located deeper in the GS model, and the
maximum abundance of this molecule is two orders of magnitude higher
than in the A5 model. A similar behavior is characteristic to the
abundances and column densities of N$_{2}$H$^{+}$, NH$_{3}$, and
H$_{2}$O (see Fig.~\ref{first}, fourth row). While in model~A5 the
column densities of these species decrease towards the inner disk,
in the models with grain growth (A4 and GS) they are nearly
constant. Again, this is a manifestation of the faster grain
evolution in the disk interior with respect to the outer region ($R
\ge 50-100$~AU).

In the inner disk midplane, where temperatures are low enough to
allow icy mantle formation, all carbon atoms are locked in surface
CO$_2$. Midplane surface and gas-phase abundances of CO and CO$_2$
do not differ in models GS and A5. At the same time, in model A5
surface hydrogenation products of these species are significantly
more abundant. For example, s-H$_2$CO and s-HCOOH (the prefix ``s-''
denotes surface species) are enhanced by one to two orders of
magnitude in the model with ``standard'' dust properties.

Another noticeable difference between models GS and A5 is the layer
of enhanced depletion located right beneath the molecular layer,
which is clearly observed in the A5 model, and is absent (or at
least less pronounced) in model GS. The origin of this depletion
layer is related to hydrogen and helium ionization. This effect is
most directly seen for CO. In the depletion layer, this molecule is
rapidly destroyed in reactions with He$^+$. Below this layer, helium
is not ionized as X-rays do not penetrate there and this channel is
not important. Above it, the reaction of CO with ionized helium
cannot compete with rich ion-molecular chemistry. In the GS model,
molecular freezing is less effective because of decreased available
dust surface, and photodesorption is enhanced; thus formation of
this depletion layer is suppressed. It is also important to note
that cosmic ray (CR) and X-ray ionization is not affected by dust
sedimentation in our model.

Abundant gas-phase water in the GS model consumes hydrogen ions
effectively (produced by CR ionization and in reactions of H$_2$
with He$^+$), and their abundance drops significantly at
intermediate heights (about 1.5 AU at $R=10$~AU). Correspondingly,
abundances grow for other species that would otherwise have been
destroyed by H$^+$. Among these species are some long carbon chains,
like C$_6$H$_2$, and cyanopolyynes, like HCN, HC$_3$N and HC$_5$N
(Fig.~\ref{second}, first and second rows). Conditions for the
gas-phase synthesis of these compounds are only met in the narrow
disk layer, and so the presence or absence of H$^+$ in this layer
determines whether or not molecules with long carbon chains can be
abundant in the inner disk. This is why their column densities are
4--5 orders of magnitude higher in the GS model, where they bind a
significant fraction of C atoms, than in the A5 model.

Thus, the typical net effect of the dust evolution
is to increase the gas-phase abundances (despite enhanced
photodissociation) of most molecules, and to decrease abundances of
their surface counterparts. Of course, given the intricate structure
of chemical networks, the response is not always straightforward,
and there are species for which {\em both\/} gas-phase and surface
abundances are increased in the GS model (like H$_2$S), and also
species for which both gas-phase and surface abundances are
decreased in the GS model (like HNO). Also, dust evolution suppresses
surface hydrogenation and, in particular, formation of the simplest
organic molecules in icy mantles.

\subsubsection{Outer disk: models GS and A4}
Yet another comparison, that is useful, is to compare the results of
models GS and A4. An increased average grain size (as well as an
increased upper limit of the grain size distribution) can be thought
of as a simplified way of modeling grain growth. The main
assumption in this representation is that dust evolves uniformly
over the entire disk. As we have seen, this is not the case, so the largest deviations should be
in those disk regions that are
less affected by grain growth and sedimentation, (in regions
further away from the star). Indeed, at a radius of 10~AU, the vertical abundance
profiles in the GS and A4 models for most species are nearly identical,
with most differences observed at $R>100$~AU.

To illustrate features specific to model A4, we turn to the
molecular disk content at 550~AU. The biggest differences in column
densities are found for surface species, which is expected, as the
grain surface area is much smaller in this model than in the other
two models. Some complex species, like s-C$_2$O and s-C$_3$O, are
underabundant in terms of column densities by more than 5 orders of
magnitude in model A4. The column densities of other carbon-bearing
surface species, like s-CO$_2$, formaldehyde, methanol, are 2--4
orders of magnitude smaller than in other models. On the other hand,
the column densities of ices, which are not affected that much by
surface chemistry (s-CO, s-H$_2$O, s-NH$_3$), differ only slightly
in models GS and A4. This is largely because these molecules simply
bind nearly all CNO atoms in both of these models.

For the gas-phase species, the largest difference in column
densities is observed for N$_2$O. This molecule is overabundant (in
terms of column densities) in model A4 relative to model GS by four
orders of magnitude. Similar overabundance is also seen for other
N-bearing species as well, like NO$_2$, OCN, and ammonia. While the
network of chemical pathways connecting all these species is quite
complicated, the underlying reason for their higher abundance seems
to be simple. Surface transformations, which are less effective at
this radius in model A4 than in model GS, lock somewhat fewer
nitrogen atoms in surface ammonia making them available for various
gas-phase species. It must be kept in mind that the column density
of s-NH$_3$ is so high that even minor relative differences in
$N$(s-NH$_3$) between models A4 and GS significantly affects column
densities of other, less abundant N-bearing species. Similar trends
are observed in other families as well. In general, we can say that
dust acts as an irreversible (or nearly irreversible) sink for
atoms, locking them in surface species with high desorption energy.
As grains get bigger, their total surface area diminishes, and more
atoms are available for richer gas-phase chemistry.

Column densities alone are not the unique indicator of a specific
chemical structure. Even if the column densities of some molecules
are almost the same in models GS and A4, their vertical
distributions can be different in terms of both width and location
of the molecular layer. In general, as we move from model A5 to GS
to A4, at large radii, the molecular layer gets thinner, and is
located deeper in the disk.

\subsubsection{Comparison to other works}

\begin{figure}
\includegraphics[width=0.9\textwidth,angle=0]{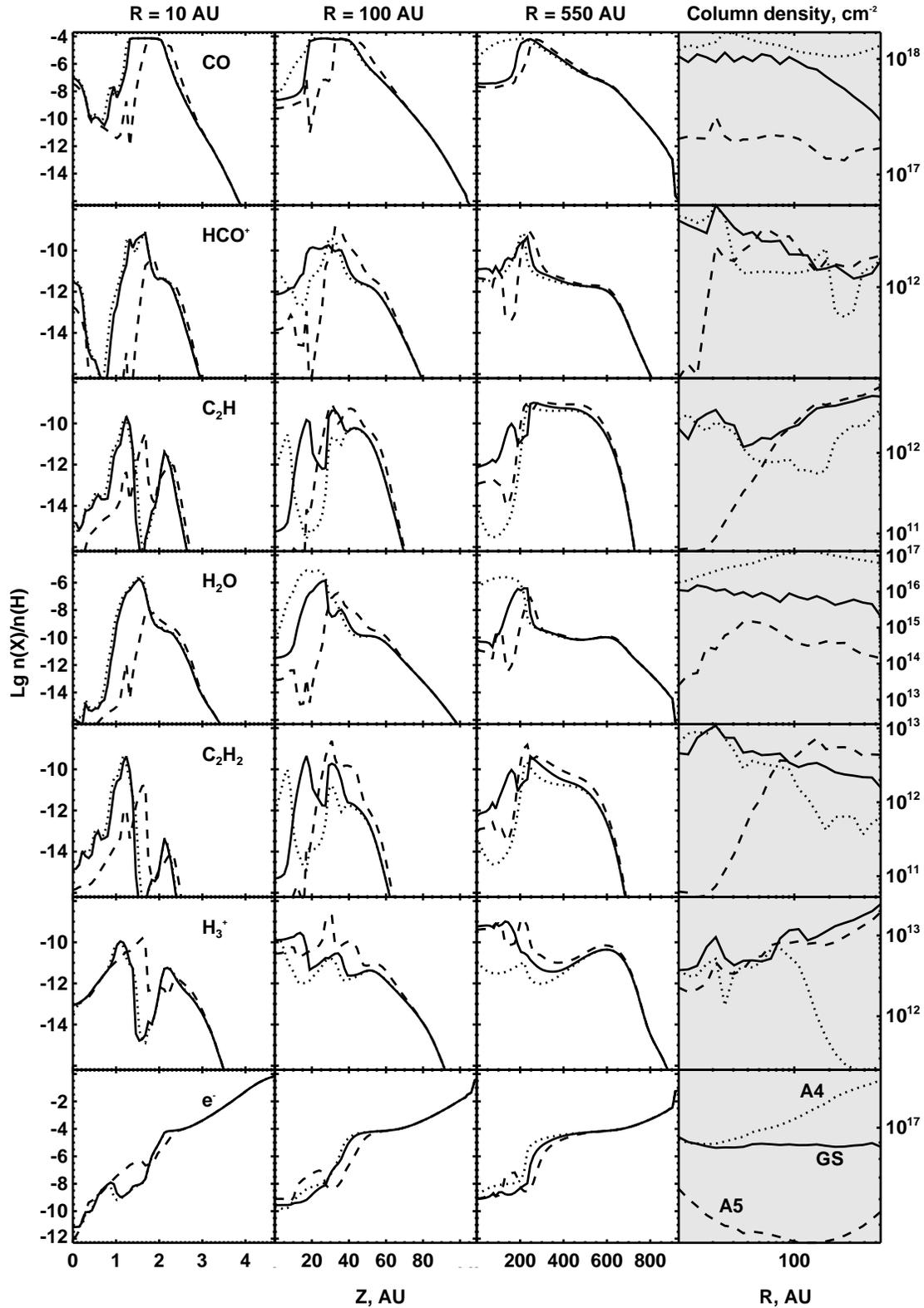}
\caption{Vertical distributions at 10, 100, and 550 AU and column
densities of CO, HCO$^+$, C$_2$H, H$_2$O, C$_2$H$_2$, H$_3^+$, and
electrons in models A5 (dashed line), GS (solid line), and A4
(dotted line).}\label{first}
\end{figure}

\begin{figure}
\includegraphics[width=0.9\textwidth,angle=0]{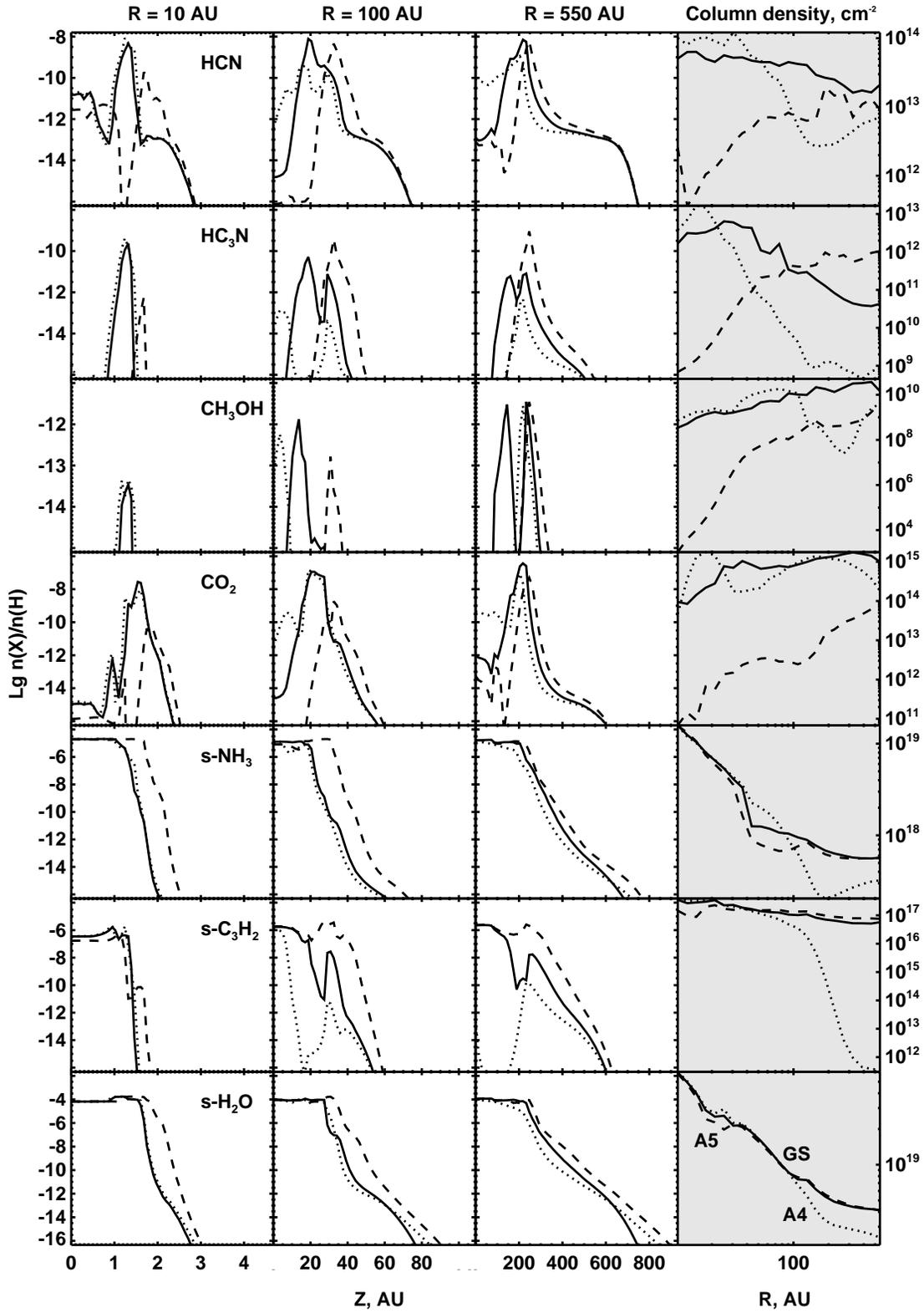}
\caption{Same as in Fig.~\ref{first}, but for HCN, HC$_3$N,
CH$_{3}$OH, CO$_2$, s-NH$_3$, s-C$_3$H$_2$, s-H$_2$O.}\label{second}
\end{figure}

The chemistry in a disk with evolved dust was studied by
\citet{AikawaNomura06}. In their investigation, H$_{3}^{+}$ was
indicated as one of a few species sensitive to the grain growth in
terms of a column density. In our study, model~A4 is most suitable
for the comparison with the model of \citet{AikawaNomura06}, and
comparison of models A5 and A4 indeed confirms the statement made in
their paper. However, H$_{3}^{+}$ behaves differently in our two
evolved models (GS and A4; see Fig.~\ref{first}, sixth row). In
model GS the column density of H$_{3}^{+}$ does not significantly
differ from that in model A5. As the formation of H$_{3}^{+}$ is
driven by cosmic ray ionization of H$_2$, with the same rate in all
three models, the difference is related to the destruction of this
ion. It is consumed mainly in dissociative recombination reactions
with electrons and negatively charged grains as well as in
ion-molecular reactions with CO, H$_2$O, CN, and other radicals.
Therefore, the distribution of H$_3^+$ in the disk is controlled by
fractional ionization and by the width and location of the molecular
layer. In models GS and A5, peaks of molecular abundances are
located at different heights, but still above the midplane. Because
of this, the vertical profiles of the H$_3^+$ abundance in these two
models are similar, except for a shift relative to the midplane.
Consequently, the {\em column density\/} of H$_3^+$ is not sensitive
to grain evolution as described by model GS.

On the other hand, in model A4 the molecular layer extends down to
the midplane, and, thus, protonation reactions involving H$_{3}^{+}$
are active in a broader disk zone. The main channels of H$_3^+$
destruction in the outer disk are its reactions with CO and water,
which are abundant even in the cold gas in model A4. This emphasizes
the need of careful grain growth modeling in chemical studies, as an
adopted approach to the grain evolution may affect conclusions even
about some primary species.

The increase in many molecular column densities seen in model GS
seems to be in contradiction with the conclusion of
\citet{AikawaNomura06} that grain growth does not affect most column
densities. However, they do not consider dust settling and also
model dust growth by increasing the upper size limit, thus, keeping
the significant amount of small grains. This means that the upper
disk in their model is, probably, more opaque than our disk in model
GS. Thus, the increase in column densities is only observed when
dust growth and settling are considered simultaneously.

Our basic results compare favorably with those obtained by
\cite{Jonkheid_ea04, Jonkheid_ea07}. However, the two models are
also quite different. In particular, freeze-out and surface
processes are not taken into account by Jonkheid et al., and also
assumptions on the central star are very different from our work.

Nevertheless, the CO depletion layer is seen in both the T~Tauri
disk model and especially in the Herbig Ae/Be disk
model, when dust is assumed to be well-mixed with the gas.
As in our model, this layer disappears when dust
settling is included in the modeling by Jonkheid et al. In both
models, the peaks in the vertical distributions of molecular
abundances occur closer to the midplane in the models with
dust evolution.

In terms of column densities, however, the two models differ. In our
model, as grains become larger, the molecular column densities
become larger. The opposite trend is observed in the model by
\cite{Jonkheid_ea07} (cf. their Fig.~8). But this difference is
understandable: the disk in their model is warmer, because the
central star is hotter and more luminous. As the disk becomes more
transparent to UV radiation, molecular column densities drop. In our
model, chemical interactions between gas and dust are very
important, and the difference between column densities in different
dust models is not controlled solely by photodissociation.

Another study to be compared with our results is those of
\cite{Pascucci_ea09}. These authors found that there is a
significant underabundance of HCN relative to C$_2$H$_2$ in disks
around cool stars and substellar objects. They attribute this
underabundance to the differences in UV radiation field. The
photodissociation of molecular nitrogen is less efficient in the
vicinity of cool stars, so there are not enough free nitrogen atoms
to be incorporated into HCN molecules, while N atoms are abundant
around hotter stars with noticeable UV excess. The UV spectrum of
the central object is the same both in GS and A5 models, however one
of the main results of grain evolution is that the integrated UV
field in the upper disk is stronger in model GS than in model A5,
and we may expect a similar trend in the HCN abundance.

And we do see this trend in the outer disk (Fig. \ref{first}, fifths
row and Fig. \ref{second}, first row). While in model A5 at
$R>100$~AU column densities of HCN and C$_2$H$_2$ are almost equal,
in model GS (with stronger UV field) HCN is 15 times more abundant
than acetylene. Analysis of chemical reactions shows that enhanced
HCN is indeed caused by greater abundance of N atoms. The trend is
not preserved at smaller radii, though. At $R=10$~AU in model A5
the column density of HCN exceeds that of acetylene by a factor of 42
(compared to near equality at greater distances). This is because in
model A5, with abundant dust in the upper disk, surface HCN
synthesis with subsequent desorption becomes an important factor
increasing gas-phase abundance of this molecule.

One should keep in mind that both under- and overabundance of HCN
relative to C$_2$H$_2$ in model GS is observed on top of the overall
trend for column densities of most molecules to be greater in this
model.

\subsubsection{Fractional ionization in the disk}

Dust evolution is reflected not only in the abundances of some
molecules, but also shows up in a more integral way, affecting the
fractional ionization $x_{\rm e}$ of the disk (see Fig. \ref{first},
bottom row). Since in our model one grain can adsorb not more than
one electron, change of dust properties impacts disk ionization
mainly through change of absorption of the UV field which is
important for gas-phase chemistry. While the ionization degree
exceeds 10$^{-10}$ everywhere in the disk, grain number density is
not more than 10$^{-12}$ in respect to hydrogen nuclei. Therefore,
ionization degree is controlled by cation chemistry while the
contribution of electron sticking to dust grains is less than 1\%.
In the disk atmosphere differences are small as the entire ion
content is provided by ionized carbon and ionized hydrogen (in the
uppermost X-ray ionized layer). Deeper in the disk, different
properties of dust cause $x_{\rm e}$ in models A5 and GS to differ
by almost two orders of magnitude. This difference is primarily
caused by different dominant ions at different heights. Let's
consider a slice at $R=100$~AU as an example. When absorption gets
high enough so that C$^+$ is no more a dominant ion ($z\la40$~AU),
in model GS most electrons are provided by metals. Sodium,
magnesium, and iron ions stay dominant down to $z\sim15$~AU. Below
this level, where most dust is concentrated in model GS, metals are
depleted, and most abundant ions are H$^+$ and H$_3^+$ (in almost
equal shares).

In model A5 dust is equally abundant everywhere in the disk, and
metals are depleted at all heights below 40~AU. Correspondingly,
almost all ions at these heights are provided by H$^+$ and H$_3^+$.
These species are heavily involved in ion-molecular chemistry, so
their abundances are very sensitive to the overall molecular
content. In particular, ionized hydrogen is effectively consumed in
reactions with ammonia, water, hydroxyl etc. These species are
abundant in the warm molecular layer ($z\sim30$~AU) and depleted
right below it ($z\sim20$~AU). Correspondingly, in the warm
molecular layer H$_3^+$ is the most abundant ion, while at lower
heights the balance shifts toward H$^+$. The equilibrium fractional
ionization (for a single ion) is inversely proportional to the square
root of the recombination coefficient. This coefficient is
significantly lower for H$^+$ than for H$_3^+$, so the fractional
ionization is higher at $z\sim20$~AU, where H$^+$ dominates.

\section{Discussion}\label{discussion}

\subsection{Disk chemistry as a tracer of grain evolution}

Can molecular abundances in a protoplanetary disk be tracers of
grain evolution? Since the distance to the Taurus-Auriga complex,
where most known disks around low-mass pre-main sequence stars are
located, is $\sim$~140~pc \citep{Elias78} and the expected angular
resolution of ALMA is up to $0".01$, the spatial resolution of disk
observations should be $\sim$~1--2~AU. This is sufficient to study
the spatial distribution of molecular abundances in the outer disk
(10--600~AU). So it is worthwhile to ask if there are species,
sensitive to the grain growth at a level that would make them
detectable tracers of this process. An immediate product of
observations are molecular lines. Therefore, we need to predict not
abundances or column densities, but the shapes and intensities of
molecular lines, as well as the variations caused by grain evolution
\citep[e.g.,][]{Semenov_ea08}. The physical model of the disk
employed in this study is quite simple, however, so it does not make
sense to perform realistic radiative transfer modeling. In this
situation, we have chosen to consider only given species as a
possible grain evolution tracer if its column density differs in
models A5 and GS by an order of magnitude or more in at least one of
representative radial regions around 10~AU, 100~AU and 550~AU. An
exception is made for CO, which has a very high column density. In
the inner disk and at R~=~100~AU from the star CO column density
becomes higher by a factor of 5--7 in model GS in comparison to
model A5.

Major species, both sensitive and insensitive to grain evolution in
our model, are grouped in Tables \ref{tab:sens_spec} and
\ref{tab:NOTsens_spec} respectively. Additionally,
Table~\ref{tab:sens_spec} shows the observed column densities for
several species, taken from \cite{Semenov_ea04} (except for
HCO$^{+}$ and N$_{2}$H$^{+}$ which are taken from Table~3 in
\cite{Dutrey_ea07} and C$_{2}$H which is taken from
\cite{Henning_ea10}). It is clear that the column densities of
species which have already been observed in disks so far, do not
seem to be strongly influenced by the changing dust properties at
large distances from the star, i.e., in the domain of single-dish
observations. Closer to the star, however, the differences are more
significant. Specifically, at R=10~AU, the column density of water
in model GS exceeds that in model A5 by a factor of 560. A word of
caution is to be said about water abundance. While column densities
for all other molecules have reasonable values in our models (as
compared to available observations), water seems to be an exception.
Recent Herschel results \citep{berginwater} show no observable water
lines in the DM Tau spectrum. \citet{berginwater} estimated the
upper limit for the water column density in the outer disk to be
about $3\cdot10^{13}$~cm$^{-2}$, which is much lower than the water
column density in our models. They suggested that low water
abundance may serve as an indication of grain growth and
sedimentation. However, as we said previously, our model do not
predict significant grain evolution in the outer disk, so the DM Tau
disk may represent a later stage of the grain growth than our model.

Of course, not only abundances (i.e., column densities)
themselves, but also their ratios, can be an indicator of the grain
growth. However, as in the example of HCN and C$_2$H$_2$, considered
above, and motivated by observations by \cite{Pascucci_ea09}, the
influence of the grain evolution on the abundance ratio can be at
least twofold, and calls for a more detailed consideration.

Another pair of observed species that may potentially be useful as a
tracer of the grain evolution is HCO$^+$ and N$_2$H$^+$
\citep{Dutrey_ea07}. While the average column densities of these
species in model GS agree quite well with observations of
\cite{Dutrey_ea07}, the ratios of their column densities in models
GS and A4 are quite different: at $R=550$~AU the ratio $N({\rm
HCO}^+)/ N({\rm N}_2{\rm H}^+)$ is 150 in model GS, and 1500 in
model A4. While the main focus of our work was on comparison between
models GS and A5, not between models GS and A4, the difference
mentioned here shows that this ratio is worth checking in a more
detailed model.

At the disk periphery,
where dust evolution is least prominent, the only species that is sensitive to
dust growth and sedimentation on the level of an order of magnitude is methanol.
However, the methanol column density is quite modest ($2\cdot10^{11}$ cm$^{-2}$)
and even smaller closer to the star.

In general, we may conclude that the chemical signatures of early grain evolution
are hardly observable with single-dish instruments, although they are a promising
target for highly sensitive interferometers such as ALMA.

The composition of the icy mantles of dust grains in the disk also may provide information about the
grain evolution process. \cite{Terada_ea07} showed that it is possible to assess the mantle
composition through observations of scattered light in edge-on disks around T~Tau stars. They found
deep water-ice absorption lines corresponding to an optical depth of $\tau$=1 in the directions of
edge-on disks HK~Tau~B and HV~Tau~C. In order to check whether ices in the observed
region are sensitive to the grain evolution process, we calculated the mantle composition in models A5 and GS
near the optical disk surface $\tau=1$ (see Table \ref{tab:ice_comp}). Due to the different dust properties in models
A5 and GS, the location of this surface is different in the two models. It is closer to the midplane in model GS,
where the disk is more transparent to UV radiation than in model A5, where the dust is pristine and more
opaque. The vertical temperature gradient of the dust
is the most important factor
in determining the differences in the ice composition between the two models. Since our model of the vertical structure of the disk is not self-consistent, the temperature at the $\tau=1$ surface  is determined only approximately.
Therefore, differences in the chemical composition of ices in models A5 and GS should only be considered as qualitative in nature.

In the inner disk ($R=10$~AU), the temperature at its optical
surface is 50~K in model GS, and 80~K in model A5. At this
temperature range, the ice composition is totally dominated by water
($\sim~99$\%). Ammonia is the second most abundant ice compound
($\sim~1$\%), and other species are present in ice only in trace
amounts. At a disk radius of 100~AU, the ice composition at $\tau=1$
is significantly different in models A5 and GS and relatively
complex. While water is still the dominant ice constituent, it
shares this role with hydrocarbons C$_{n}$H$_{m}$, which are
produced efficiently by both surface chemistry and gas-phase
chemistry in this disk region. Also, in the GS model, several other
species, such as CO$_{2}$, H$_{2}$CO, HCN etc. stay in ice in
noticeable amounts. Further away from the star, at a radius of
550~AU, the ice composition in models A5 and GS is almost the same
again because the optical surface at this distance in both models
corresponds to very similar temperatures of approximately 30~K.

Such differences in grain mantle composition at the optical surface of the disk may serve as an additional tracer
of the grain evolution process. However, the results described above should be confirmed with detailed,
self-consistent models of the vertical structure of the disk.

\subsection{Model limitations and future work}

We have already mentioned that column densities are only a partial indicator
of grain growth. Line emission modeling is needed to assess observable
tracers of dust evolution with more confidence. But, there are some issues
that need to be addressed prior to this modeling. As we mentioned in the
introduction, the simplified setup of our model allows us to isolate the chemical
outcomes of grain evolution, although the process of grain growth should probably
have some more general influence on the disk structure. The response
of a disk to dust evolution has already been addressed a number of times but
with only an oversimplified approach to grain growth and/or sedimentation.
Our modeling shows that dust evolution in a disk is highly non-uniform,
both in the vertical and radial direction, which should be reflected in disk physical properties.

In future disk models, one would need to at least calculate the
processes of grain growth and sedimentation consistently with the
description of the disk structure, as density and temperature
distributions in the disk are controlled by opacities. Dust
absorption is definitely affected by growth and sedimentation, but
both these processes, in turn, depend on the disk structure. In this
study, changes in the surface-averaged grain size (and, hence, in
the total grain surface area) are only minor, and even in model GS,
where dust is evolved, the disk in relatively opaque to the incoming
radiation. This means that extra heating due to decreased opacity
would mostly affect the upper disk, above the region where most
molecules are concentrated. As in this region molecular abundances
are mostly controlled by photoprocesses, most column densities
probably will not be strongly affected by the increased gas scale
height or gas temperature in the upper disk. The exception is
represented by abundant components that bind almost all CNO atoms in
the molecular layer (CO, N$_2$, O). As we noted, the increase of
their column densities in the inner disk in models GS and A4
relative to A5 is partially caused by the downward shift of the
molecular layer to the denser disk region. If we would take the disk
vertical expansion due to increased heating into account, the peak
of relative abundance for these components would probably reside in
the region of smaller density, thus, suppressing the increase in
column densities. This calls into question the ability of CO to be a
tracer of dust evolution within the framework of our model. However,
dust growth may affect not CO itself, but its isotopic ratios
\citep{visser_ea09}.

Another issue, which should be addressed properly in the future, is
the possible interrelation between chemical and physical structure
of the disk with evolving dust. This interrelation has already been
discussed from the point of view of thermal balance in the disk
\citep[e.g.,][]{Jonkheid_ea07}. However, dust evolution may also
lead to significant changes in the fractional ionization, which is
an important dynamical parameter, especially if turbulence in disks
is excited due to the magnetorotational instability
\citep{IlgnerNelson08}. All of this calls for more self-consistency
not only between disk structure and grain evolution, but also
between these two factors and chemistry.

The next step is to include time dependence in the model. Two of the
three ingredients of our model (disk structure and grain ensemble
properties) explicitly imply that the disk is in a steady state. The
third ingredient (chemistry) is formally time-dependent, but
produces steady-state abundances for most species. This is the
reasonable outcome, given the equilibrium nature of the other two
ingredients. Our results represent only a snapshot of the disk at an
early (or even the earliest) stage of grain evolution. However, at
some later stage, the grain growth timescale may become short enough
to be comparable with the timescale of various chemical
transformations. With the current model we cannot say what impact on
chemistry possible further grain growth, beyond what the Birnstiel
model can currently calculate, would have, even though some hints
are provided by model A4. In particular, we may expect higher
molecular densities at the disk periphery, with the molecular layer
extending down to the midplane. This confirms the earlier suggestion
that observations of gas-phase CO and water molecules in a cold disk
region can be explained by an increase in grain size. Further
improvements in the dust growth model are also needed. This field
now represents an area of active research, both theoretical and
experimental, and there are certain indications that grain evolution
is not limited to simply sticking and fragmenting
\citep[e.g.,][]{Zsom_ea10}.

Another issue that may become more important with the grain ensemble
containing large grains is the treatment of grain surface chemistry.
Currently, in chemical modeling we consider grains of equal
``effective mean size" instead of the real grain size distribution.
We believe that this simplification does not dramatically affect
results of this study. The average grain size defined according to
(\ref{meanrad0}) is close to the size of grains that dominate the
total grain surface area available for surface chemistry. In other
words, most of the surface chemistry occurs on grains of similar
sizes. Also, Acharyya \& Herbst (in preparation) constructed a
chemical model with grains of several grain sizes in order to
reproduce the grain size distribution. They found its results to be
quite similar to those obtained with a single grain size. However,
the problem may be more complicated as the temperature of an
individual grain depends on its size. For smallest grains,
stochastic heating becomes important, which makes it impossible to
assign a single temperature to them. These questions are not trivial
and worth to be considered in a separate study. In this work, we do
not have grains of extremely small radii ($R\le10^{-6}$~cm). So, the
considerations above should be of only a limited importance here.

Finally, transport processes are an important evolutionary factor in an
{\em accretion\/} disk. At least, some of the differences between models GS and A5, like the enhanced depletion layer in model A5, are quite localized.
One does need to account for possible gas mixing to check whether these
differences can survive in a more dynamic medium. The radial motion of grains has to be included in a realistic treatment of the disk evolution.

\section{Conclusions}\label{conclusions}

We investigated the chemical evolution of the protoplanetary disk
around a T Tauri star. The thermal and density structures of the
disk are calculated with a 1+1D $\alpha$-viscosity model. The
processes of grain coagulation, fragmentation, and sedimentation
were modeled as described in \citet{Brauer_ea08} and
\citet{Birnstiel_ea10}. The initial dust ensemble is the standard
MRN grain size distribution \citep{Mathis_ea77}. Using the final
grain size distribution (at $t=2$~Myrs) and the 1+1D disk structure,
we calculated the vertical distribution of the dust in the disk,
assuming stirring-settling equilibrium according to
\citet{DullemondDominik04}. For this distribution, we calculated the
approximate UV field in the disk, and simulated the chemical
evolution of the disk over 2~Myr. Results for three dust models were
compared: (1) a classical dust model, with uniform spherical
particles of $0.1\mu$m and a gas-to-dust mass ratio of 100
(model~A5); (2) a simplified grain growth model, with larger grains
of $1\mu$m (model~A4); (3) a detailed grain growth model (model~GS).
The main results of the work can be summarized as follows:

\begin{itemize}

\item
The fragmentation of grains due to collisions keeps a population of small dust particles
in the disk, which dominate the total grain
surface area crucial for chemical evolution of the disk. This effect reduces the impact of grain growth
on the chemical structure of the disk.
Grain growth proceeds fast in low-mass T Tauri disks at distances $\le 20-50$~AU where the steady-state grain size
distribution is reached in a few thousand years.
Outside this radius, up to 10$^{6}$~years is needed to reach steady state at R$\ge$100~AU.
Furthermore, grain coagulation
is limited there, and increases the average grain size by
no more than an order of magnitude in the midplane.
After 2~Myrs of evolution, the average grain size in the disk midplane varies
between $10^{-4}$~cm at $R=10$~AU, and $2\cdot 10^{-5}$~cm at
$R=550$~AU. The dust-to-gas mass ratios in the same distance range are confined between $\sim 10^{-6}$
and $0.02$.

\item
The net effect of the dust settling and growth is a reduction of
the total grain surface, and higher UV irradiation rates in the upper disk. The chemical
structure of an evolved disk still has three layers, but the
intermediate molecular layer gets wider, and shifts closer to the
midplane. Therefore, the abundances and column densities of many species are
enhanced by a factor of
$3-100$, even for such a moderate grain growth.

\item
A simplified model of grain growth, in which the dust-to-gas mass
ratio is kept constant and grain size is simply increased by one order of magnitude, is not sufficient to
reproduce the chemical evolution of a disk with evolving dust. For
example, the column density of H$_{3}^{+}$, which is proposed to be
sensitive to grain growth \citep{AikawaNomura06},
exhibits a high sensitivity to grain growth in our simple model, but not in
the model with a detailed treatment of grain growth and sedimentation.

\item Grain evolution suppresses the formation of organic molecules in icy mantles, but
also enhances gas-phase molecular abundances.
In particular, water and CO abundances are enchanced in the cold midplane.

\item
We propose a few observational tracers of grain evolution process.
These are the column densities of gas-phase molecules, such as
C$_{2}$H, HC$_{2n+1}$N ($n=0-3$), H$_{2}$O for the inner disk. The
abundance ratio of C$_{2}$H$_{2}$/HCN may serve as a grain evolution
tracer in outer disk, as well as the abundance of CH$_{3}$OH. The
composition of the ice mantles of interstellar grains seen in
scattered light in the intermediate disk layer (around $\tau$=1) may
also serve as a tracer of grain evolution.

\end{itemize}

\section{Acknowledgements}

We are grateful to Dr. Dmitry Semenov for fruitful and stimulating
discussions, to Prof. J\"urgen Blum for the information about
collisional properties of different dust materials and to anonymous
referee whose suggestions helped to improve the presentation of our
results. DW's work is supported by the Federal Targeted Program
``Scientific and Educational Human Resources of Innovation-Driven
Russia'' for 2009-2013. This research has made use of NASA's
Astrophysics Data System.

\bibliographystyle{apj}

\clearpage

\begin{table}
\caption{Main disk model parameters}
\label{modpar}
\begin{tabular}{l|l}
\hline\hline
Inner radius & 0.03 AU \\
Outer radius & 550 AU \\
$\alpha$ & 0.01 \\
$\dot{M}$, $M_\odot$ year$^{-1}$ & $2\cdot10^{-9}$ \\
X-ray luminosity & $10^{30}$ erg s$^{-1}$ \\
Cosmic ray ionization rate & $1.3\cdot10^{-17}$ s$^{-1}$ \\
Disk mass & 0.055 $M_\odot$\\
Initial average grain size & 10$^{-5}$~cm \\
\hline\hline
\end{tabular}
\end{table}

\clearpage

\begin{landscape}

\begin{table}
\caption{Species sensitive to grain evolution. Observed column
densities are compiled from \cite{Dutrey_ea97, Qi00, Aikawa_ea02,
Dutrey_ea07}, \cite{berginwater} and \cite{Henning_ea10}.}
\label{tab:sens_spec} \centering
\begin{tabular}{l|cccccc|cc|cc}
\hline \hline
Species  & \multicolumn{6}{|c|}{Column densities, cm$^{-2}$} & \multicolumn{2}{|c}{Peak abundance} & \multicolumn{2}{|c}{Observed} \\
         & \multicolumn{2}{c}{10 AU} & \multicolumn{2}{c}{100 AU} & \multicolumn{2}{c|}{550 AU} & \multicolumn{2}{c}{n(X)/n(H)} & \multicolumn{2}{|c}{column densities, cm$^{-2}$}\\
         & A5 & GS & A5 & GS & A5 & GS & A5 & GS & DM Tau & LkCa15 \\
\hline
              CO & 2.0(17) & 1.1(18) & 2.0(17) & 9.4(17) & 1.7(17) & 2.9(17) & 7.3(-05) & 7.3(-05) & 5.7(16) & 9.0(17) \\
        CO$_{2}$ & 6.8(10) & 9.5(13) & 2.4(12) & 8.5(14) & 8.2(13) & 9.8(14) & 7.2(-08) & 4.6(-07) & - & - \\
      CH$_{3}$OH & 1.1(03) & 3.5(08) & 1.9(08) & 1.0(10) & 4.6(09) & 1.4(10) & 1.2(-11) & 1.2(-11) & - & - \\
        H$_{2}$O & 2.5(13) & 1.1(16) & 3.6(14) & 7.0(15) & 1.4(14) & 2.0(15) & 1.3(-06) & 2.6(-06) & $\le$3.0(13) & - \\
        H$_{2}$S & 1.7(05) & 1.3(10) & 2.0(09) & 1.1(11) & 2.9(10) & 3.6(10) & 2.0(-11) & 1.9(-11) & - & - \\
        C$_{2}$H & 6.3(10) & 2.0(12) & 2.6(12) & 2.2(12) & 6.7(12) & 5.0(12) & 1.4(-09) & 1.5(-09) & 2.8(13) & 2.9(13) \\
  C$_{2}$H$_{2}$ & 6.2(10) & 4.6(12) & 4.5(12) & 3.0(12) & 4.4(12) & 1.6(12) & 4.0(-09) & 1.3(-09) & - & - \\
CH$_{3}$CH$_{3}$ & 3.0(09) & 1.5(12) & 1.8(10) & 3.3(08) & 1.3(07) & 2.5(06) & 3.6(-11) & 2.3(-10) & - & - \\
       H$_{2}$CS & 8.0(05) & 3.5(11) & 2.7(11) & 8.3(11) & 2.7(10) & 1.8(11) & 2.2(-10) & 5.6(-10) & - & -\\
             HCN & 2.6(12) & 5.0(13) & 6.9(12) & 4.2(13) & 9.0(12) & 2.1(13) & 1.8(-08) & 9.1(-09) & 2.1(12) & 7.8(13) \\
       HC$_{3}$N & 6.8(08) & 1.7(12) & 4.1(11) & 2.8(11) & 1.1(12) & 4.3(10) & 9.1(-10) & 1.1(-09) & - & - \\
       HC$_{5}$N & 4.9(08) & 1.5(12) & 2.1(11) & 1.5(11) & 1.8(11) & 8.8(10) & 5.9(-10) & 6.5(-10) & - & - \\
       HC$_{7}$N & 4.7(06) & 1.2(12) & 6.1(10) & 5.3(10) & 9.2(10) & 2.8(10) & 3.9(-10) & 6.0(-01) & - & - \\
      HCNH$^{+}$ & 2.0(10) & 2.7(11) & 5.7(10) & 1.3(11) & 6.6(10) & 6.8(10) & 1.1(-10) & 3.6(-11) & - & - \\
           HCOOH & 1.4(11) & 6.3(13) & 1.1(12) & 1.3(13) & 2.6(11) & 2.1(12) & 4.0(-09) & 1.2(-08) & - & - \\
             OCN & 3.2(07) & 4.3(09) & 2.3(10) & 1.5(12) & 1.3(11) & 4.2(12) & 2.8(-10) & 4.3(-09) & - & - \\
             OCS & 1.6(07) & 9.3(10) & 1.6(10) & 1.5(10) & 2.7(08) & 4.5(08) & 1.9(-11) & 1.9(-11) & - & $\le$2.9(13) \\
        NH$_{3}$ & 1.7(11) & 3.2(13) & 1.1(13) & 3.5(13) & 8.5(12) & 1.2(13) & 1.3(-08) & 9.4(-09) & - & - \\
       HCO$^{+}$ & 1.1(11) & 5.2(12) & 2.6(12) & 2.2(12) & 2.2(12) & 1.9(12) & 2.0(-09) & 7.0(-10) & 6.5(12) &  8.0(12) \\
              OH & 6.1(13) & 1.9(14) & 2.7(13) & 1.2(14) & 1.9(13) & 1.5(14) & 3.4(-08) & 7.0(-08) & - & - \\
\hline \hline
\end{tabular}
\end{table}

\end{landscape}

\clearpage

\begin{landscape}

\begin{table}
\caption{Species insensitive to grain evolution. Observed column
densities are compiled from \cite{Dutrey_ea97, Qi00} and
\cite{Dutrey_ea07}.} \label{tab:NOTsens_spec} \centering
\begin{tabular}{l|cccccc|cc|cc}
\hline \hline
Species  & \multicolumn{6}{|c|}{Column densities} & \multicolumn{2}{|c}{Peak abundance} & \multicolumn{2}{|c}{Observed} \\
         & \multicolumn{2}{c}{10 AU} & \multicolumn{2}{c}{100 AU} & \multicolumn{2}{c|}{550 AU} & & & \multicolumn{2}{|c}{column densities, cm$^{-2}$}\\
         & A5 & GS & A5 & GS & A5 & GS & A5 & GS & DM Tau & LkCa15 \\
\hline
            CN & 6.8(12) & 2.9(13) & 5.6(12) & 9.6(12) & 1.2(13) & 1.7(13) & 3.6(-09) & 2.7(-09) & 9.5-12(12) &  6.3(14) \\
             C & 5.1(16) & 5.9(15) & 6.3(16) & 7.4(16) & 1.6(17) & 2.1(17) & 5.6(-05) & 5.4(-05) & - & - \\
   H$_{3}^{+}$ & 2.3(12) & 3.7(12) & 8.5(12) & 1.2(13) & 2.0(13) & 2.5(13) & 2.4(-09) & 1.0(-09) & - & - \\
N$_{2}$H$^{+}$ & 3.0(09) & 2.4(10) & 1.7(10) & 1.7(10) & 1.9(10) & 2.2(10) & 6.1(-11) & 2.4(-11) & 1.1(11) & 2.3(11)  \\
           HNC & 1.9(12) & 4.3(12) & 2.5(12) & 1.5(13) & 5.4(12) & 1.3(13) & 6.3(-09) & 4.7(-09) & 9.1(11) & $\le$5.4(12) \\
       C$^{+}$ & 2.5(16) & 6.2(16) & 3.3(16) & 7.8(16) & 5.0(16) & 8.2(16) & 7.3(-05) & 7.3(-05) & - & - \\
      CH$_{3}$ & 2.8(11) & 1.7(12) & 4.7(12) & 3.0(12) & 2.4(13) & 1.8(13) & 7.0(-09) & 1.7(-09) & - & - \\

\hline \hline
\end{tabular}
\end{table}

\end{landscape}

\clearpage

\begin{table}
\caption{Ice composition in the disk at the surface $\tau=1$. Only species with contribution of 1\%
or more are shown}
\label{tab:ice_comp}\centering
\begin{tabular}{l|llllll}
\hline \hline
Species & \multicolumn{6}{c}{Ice fraction, \%} \\
         & \multicolumn{2}{c}{10 AU} & \multicolumn{2}{c}{100 AU} & \multicolumn{2}{c}{550 AU} \\
         & A5 & GS & A5 & GS & A5 & GS \\
\hline
s-H$_{2}$O       & 99.8 & 97.3 & 57.4 & 91.0 & 70.0 & 88.9 \\
s-NH$_{3}$       & -    & 1.0  & -    & -    & 1.0  & 5.0  \\
s-C$_{2}$H$_{2}$ & -    & -    & -    & -    &  4.6 & 1.0  \\
s-C$_{3}$H$_{2}$ & -    & -    & 20.9 & 3.3  & 16.5 & 1.5  \\
s-C$_{5}$H$_{2}$ & -    & -    & 12.0 & -    & 2.1  & -    \\
s-C$_{7}$H$_{2}$ & -    & -    & 5.0  & -    & -    & -    \\
s-C$_{9}$H$_{2}$ & -    & -    & 1.3  & -    & -    & -    \\
s-CO$_{2}$       & -    & -    & -    & 2.0  & -    & -    \\
s-H$_{2}$O$_{2}$ & -    & -    & -    & -    & -    & 1.5  \\
s-HCN            & -    & -    & -    & -    & 3.1  & 1.0  \\
s-HNO            & -    & -    & -    & 1.0  & -    & -    \\
s-H$_{2}$C$_{4}$ & -    & -    & 1.0  & -    & 1.0  & -    \\
\hline
\end{tabular}
\end{table}

\clearpage

\clearpage

\clearpage

\clearpage

\clearpage

\clearpage

\clearpage

\clearpage

\clearpage

\clearpage

\clearpage

\end{document}